\documentclass[twocolumn,preprintnumbers,amsmath,amssymb,floatfix,prd,nofootinbib]{revtex4}
\pdfoutput=1
\usepackage{epsfig}
\usepackage{dcolumn}
\begin{document}
\title{Revisiting Helicity Parton Distributions at a Future Electron-Ion Collider}
\author{Ignacio Borsa}
\email{iborsa@df.uba.ar}
\affiliation{Departamento de F\'{\i}sica and IFIBA,  Facultad de Ciencias Exactas y Naturales, 
Universidad de Buenos Aires, Ciudad Universitaria, Pabell\'on\ 1 (1428) Buenos Aires, Argentina}
\author{Gonzalo Lucero}
\email{glucero@df.uba.ar}
\affiliation{Departamento de F\'{\i}sica and IFIBA,  Facultad de Ciencias Exactas y Naturales, 
Universidad de Buenos Aires, Ciudad Universitaria, Pabell\'on\ 1 (1428) Buenos Aires, Argentina}
\author{Rodolfo Sassot}
\email{sassot@df.uba.ar} 
\affiliation{Departamento de F\'{\i}sica and IFIBA,  Facultad de Ciencias Exactas y Naturales, 
Universidad de Buenos Aires, Ciudad Universitaria, Pabell\'on\ 1 (1428) Buenos Aires, Argentina}
\author{Elke C. Aschenauer}
\email{elke@bnl.gov}
\affiliation{Physics Department, Brookhaven National Laboratory, Upton, NY 11973, USA
}
\author{Ana S. Nunes}
\email{anunes@bnl.gov}
\affiliation{Physics Department, Brookhaven National Laboratory, Upton, NY 11973, USA
}
\begin{abstract}
We studied the impact of future Electron Ion Collider  inclusive and semi-inclusive polarized deep 
inelastic scattering data  will have on the determination of the helicity parton distributions.
Supplementing the Monte Carlo sampling variant of the DSSV14 analysis with pseudo-data on polarized inclusive and semi-inclusive electron-proton deep inelastic scattering with updated uncertainty estimates and for two different center-of-mass-system energies, $\sqrt{s}=44.7$ GeV and $\sqrt{s}=141.4$ GeV respectively, and on inclusive electron-helium collisions at $\sqrt{s}=115.2$ GeV,
we find a remarkable improvement in the determination of the helicity distributions, specially at low parton momentum fraction $x$. 
While inclusive electron-proton data at the lowest energy configuration constrain significantly the gluon polarization down to $x \sim 10^{-4}$, the higher energy configuration strengthens the constraint and extends it one decade further. On the other hand, semi-inclusive data achieves the hitherto elusive flavor separation for sea quarks that can not be obtained from any other inclusive electromagnetic measurement. Collisions with helium complement inclusive proton measurements, pushing the constraints 
on the combined quark plus anti-quark $u, d$ and $s$ polarizations to an unprecedented level. 

\end{abstract}
%
%
\maketitle

\section{Introduction and Motivation}
%
Ever since the pioneering measurements of the EMC experiment at CERN suggested that quarks and 
anti-quarks are only responsible for a small fraction of the proton spin \cite{ref:emc-a1p}, thus challenging the naive quark-parton model picture, the way in which the proton spin builds up from its fundamental constituents, quarks and gluons, has remained an open question \cite{Aidala:2012mv}.

While the amount of spin carried by quarks and anti-quarks has been confirmed to be much less than the expected $\frac{\hbar}{2}$ value by subsequent measurements of polarized Deep Inelastic Scattering (DIS) 
of leptons off fixed proton, deuteron and helium targets at SLAC, CERN, DESY and JLAB \cite{Aidala:2012mv}, it is still unclear how much of the missing spin is carried by the gluons, and how much should be associated to the orbital angular momentum of partons.  

In this quest for the origin of the proton spin, the measurement of hadrons and jets produced at high transverse momentum in polarized proton-proton collisions at BNL-RHIC has also set a crucial milestone \cite{Aschenauer:2015eha}. Since these measurements receive their most significant contributions from gluon initiated processes, the RHIC spin program has provided a fundamental grip on the gluon polarization, showing a non negligible contribution to the spin of the proton, albeit in a restricted gluon momentum fraction region\cite{deFlorian:2014yva,Nocera:2014gqa}. In spite of the very successful RHIC spin program, the gluon helicity distribution can be at best conjectured for values of the momentum fraction 
below $x \sim 10^{-2}$ . 

Besides these fundamental questions on the role of the gluon polarization, and that of the orbital angular momentum, the way in which each particular quark flavor contributes to the proton spin is also work in progress. In the naive quark model there is a fundamental relation between quark spin and flavor. However, we still lack a clear picture of how quarks of radiative origin modify at short distances the quark-parton model paradigm. First determinations of flavor discriminated helicity parton densities were based solely on rather simple symmetry assumptions. Even though these were later replaced or supplemented with Semi Inclusive Deep Inelastic Scattering (SIDIS) data, and recently with data from charged 
weak vector boson production in proton-proton collisions, the kinematical coverage for this data is  limited and therefore non conclusive \cite{Aidala:2012mv}. 

The Electron Ion Collider (EIC) is designed to be the most powerful tool to answer the above mentioned questions, quantifying the way in which gluons and quarks of different flavors make up for the total spin of the proton \cite{Accardi:2012qut}. 
Specifically, with unprecedented precision and a wide coverage in the parton momentum fraction $x$ and in the photon virtuality $Q^{2}$, the EIC is expected to provide new insights into the gluon polarization. In particular, extremely precise measurements of the polarized structure function $g_{1}$ and its scaling violations for values of $x$ two decades below the limit of $10^{-2}$ will lead to stringent constraints on the helicity distributions, absent in current global analysis. 
On the other hand, a very thorough program of SIDIS measurements (polarized and unpolarized) at the EIC will allow to develop a much more precise picture of the relations between spin and flavor, 
the role of sea quarks in the proton polarization, and the degree of flavour and charge symmetry-breaking. It has already been shown that SIDIS measurements are instrumental for discriminating between quark and antiquarks, but will be extended one step further in precision and range at an EIC. Although naively one would expect quarks to become charge and flavor symmetric as well as unpolarized at small enough parton momentum fraction, none 
of the available estimates predict where exactly QCD radiative mechanisms overcome non-perturbative effects. We note that unpolarized measurements of the SIDIS cross-sections in the same broad kinematic region where the spin dependent data is obtained are expected to allow us to further refine the current perturbative description of SIDIS processes, namely providing fragmentation functions 
of unprecedented precision \cite{Aschenauer:2019kzf}. 
Additionally, even though inclusive electron-helium collision can not 
disentangle quarks from antiquarks, very precise measurements of this type complement inclusive and semi-inclusive electron-proton data by providing access to the total (quark plus antiquark) polarization for u, d and s quarks with extreme precision down to very small parton momentum fractions. 

In order to asses the impact of future EIC data in the determination of the helicity parton distributions, we have perform a NLO global analysis adding to the DSSV14 data set \cite{deFlorian:2014yva}, inclusive and semi-inclusive polarized DIS pseudo-data for EIC with realistic uncertainties. The analysis follows the lines of those in references \cite{Aschenauer:2012ve} and \cite{Aschenauer:2015ata}, however with updated kinematics and improved uncertainty estimates. 
Different to \cite{Aschenauer:2012ve}, that relied on DSSV08 helicity distributions as theory input \cite{deFlorian:2008mr,deFlorian:2009vb}, we start from the DSSV14 results to generate the observables that are subsequently smeared according to the expected experimental errors. We also incorporate semi-inclusive estimates, that were not discussed in \cite{Aschenauer:2015ata} and study 
their impact on the quark helicity distributions. The analysis is carried out within the Monte Carlo sampling framework to estimate uncertainties, as introduced in \cite{deFlorian:2019zkl}. The method relies on the generation of a set of replicas that are assumed to give a faithful representation of the underlying probability distribution of the helicity parton densities, and helps as an alternative to the Lagrange multiplier approach used in \cite{Aschenauer:2012ve} and \cite{Aschenauer:2015ata}, that avoids the explicit adoption of tolerances, and also allows reweighting for different data sets.

Rather than to reweight the DSSV14 replicas obtained in \cite{deFlorian:2019zkl}, we generate a new set already incorporating the pseudo-data corresponding to the inclusive DIS measurements at the lowest c.m.s. energy configuration ($\sqrt{s}=45$ GeV), that presumably will be one of the first results obtained at the EIC. 
In this way, we enhance the number of surviving replicas after subsequent reweighting with other 
pseudo-data, and make the sample large enough to derive statistically meaningful results and to 
discuss and compare their specific impact.
 
As result of the above mentioned procedure we find that inclusive electron-proton data at the lowest energy configuration not only constrain significantly the gluon polarization down to $x \sim 10^{-4}$ as expected, but have also a sizable impact up to large values of $x$, thus constraining also the net polarization of the gluons in the region of momentum fractions were it is expected to get its most significant contributions. The total quark contribution $\Delta \Sigma$ is also significantly constrained compared to the original DSSV14 results. However, there is only a marginal impact on the  polarization of the sea quarks. Adding the inclusive electron-proton DIS data at the higher c.m.s. energy ($\sqrt{s}=141$ GeV) 
roughly duplicates the impact on the gluon polarization uncertainty and extends the constraints one decade further in the parton momentum  fraction. In this way the EIC would probe partons carrying down to a hundred thousandth of the proton momentum. 
Semi-inclusive data complement inclusive measurements and achieve flavor separation between quarks and antiquarks. In a similar way, electron-helium inclusive measurements complement inclusive proton ones, pushing the constraints on the total quark polarizations for the different flavors to an unprecedented level. 

The remainder of the paper is organized as follows: in the next section we discuss the generation of the different pseudo-data for the inclusive and semi-inclusive measurements used in our analysis, and their corresponding uncertainty estimates. We also present in that section an assessment of the kinematical dependence of the impact one could anticipate studying the correlation and sensitivity coefficients for the most relevant quantities.  Next, we briefly discuss the way in which we produce helicity distribution replicas from the DSSV14 data set supplemented with the low c.m.s. energy electron-proton inclusive DIS pseudo-data. In Section IV, we discuss the impact of both lower and higher c.m.s. energy inclusive pseudo-data sets and compare with the original DSSV14 estimates. We also present the impact of the
semi-inclusive data, and of the electron-helium measurements, respectively.  Finally, we summarize our results and present our conclusions.

%
\section{Pseudodata for polarized DIS and SIDIS at the EIC}\label{sec:pseudo}
\subsection{Generation of pseudodata }\label{subsec:gen}

To quantify the impact of an EIC on our understanding of helicity PDFs, we generate sets of pseudo-data for lepton proton and lepton Helium-3 scattering at center-of-mass energies of $\sqrt{s}$= 44.7 GeV, 141.4 GeV and 115.2 GeV, respectively.
We use the PEPSI Monte Carlo (MC) generator \cite{Mankiewicz:1991dp} to produce the EIC pseudo-data for
the inclusive and semi-inclusive DIS of longitudinally polarized electrons and protons with identified charged pions and kaons in the final-state. We demand a minimum $Q^2$ of $1\,\mathrm{GeV}^2$, a squared invariant mass of the virtual photon-proton system $W^2$ larger than 10 GeV$^2$, and $0.01\le y\le 0.95$.
The range of $y$ is further restricted from below by constraining the depolarization factor of the virtual photon 
\begin{equation}
\label{eq:depol}
D(y) = \frac{y(y-2)}{y^2+2(1-y)(1+R)}
\end{equation}
to be larger than $0.1$. $R$ denotes the ratio of the longitudinal to transverse virtual photon
cross sections. To ensure detection of the scattered lepton we require a minimum momentum of $0.5\,\mathrm{GeV}$, and, in case of SIDIS, only hadrons with a momentum larger than $1\,\mathrm{GeV}$ 
and a fractional energy in the range $0.2\le z\le 0.8$ are accepted.
All particles detected in the final-state in $-4 < \eta < 4$ in rapidity, corresponding to be at least 
$> 2$ degree away from the beam lines. The statistical accuracy of each DIS and SIDIS data set corresponds to a accumulated integrated luminosity of $10\,\mathrm{fb}^{-1}$.

The PEPSI MC allows one to generate events with definite helicities of the colliding lepton and proton beams, i.e., to study the longitudinal double-spin asymmetry. 
\begin{eqnarray}
\label{eq:asym}
A_{\parallel} (x,Q^2) &=& \frac{d\sigma^{++}-d\sigma^{+-}}{d\sigma^{++}+d\sigma^{+-}} \\
\label{eq:ratio}
&=& D(y) \frac{g_1(x,Q^2)}{F_1(x,Q^2)}
\end{eqnarray}
which is related to the ratio of virtual photoabsorption cross sections, expressed by DIS structure functions in (\ref{eq:ratio}), through the depolarization factor $D(y)$. In (\ref{eq:ratio}), and also in (\ref{eq:depol}), we have neglected kinematic corrections proportional to $\gamma=\sqrt{4M^2 x^2/Q^2}$, with $M$ the proton mass, which are negligible at a collider.
While containing spin-dependent hard scattering matrix elements at $\cal{O}(\alpha_{\mbox{s}})$ accuracy, the PEPSI MC is not capable of simulating parton showers which properly track the polarization of the partons involved, and hence this option has been turned off for generating the EIC data. QED radiative corrections are known to be sizable and complicate the determination of the ``true'' values of $x$ and $Q^2$. We do not consider QED radiative corrections to be a major limitation on proposed DIS and SIDIS measurements at an EIC as unfolding there effects on the measurements has become standard at lepton-hadron experiments, i.e. H1, Zeus, HERMES, COMPASS. Available MC tools \cite{ref:qed} will be further refined in the upcoming years; there is already a significant ongoing effort to implement QED radiative effects also in the general purpose generators, i.e. PYTHIA-8 \cite{Sjostrand:2014zea}.

As in Ref. \cite{Aschenauer:2015ata}, the actual pseudo-data used in  
our analyses below are not the generated ones but theoretical  
estimates of the spin asymmetries at NLO accuracy based on the latest  
DSSV helicity densities
\cite{deFlorian:2014yva} and fragmentation functions  
\cite{deFlorian:2014xna,deFlorian:2017lwf}, reflecting the same  
relative statistical accuracy in each $x,Q^{2}$ bin as the Monte Carlo  
data, and having their central values randomized within one-sigma  
uncertainties.
In our simulations of DIS and SIDIS off polarized neutrons, we assume that the experiment uses a polarized $^{3}He$ beam. To ensure that the scattering happened on the neutron one requires the spectator protons to be tagged, a commonly used technique. For EIC great care is taken to integrate detectors along the outgoing hadron beam into the wider interaction region detecting spectator protons, protons from diffractive reactions and nuclear breakup and to guarantee high detection efficiencies of $>80\%$.
The size of the asymmetry sets the scale at which one needs to control systematic uncertainties
due to detector performance or luminosity measurements. Until the EIC detector design is finalized the approach taken for the systematic uncertainties is to assume that one can reach what was achieved at HERA, for inclusive measurements a 1.6 \% and for the SIDIS measurements 3.5 \% point-to-point systematic uncertainty is assumed.  In addition a scale uncertainty due to the luminosity and lepton and hadron polarization measurement of 2.3 \% has been assigned. We have checked that adding up to a 2 \% systematic uncertainty to DIS  
pseudo-data and 3.5\% to SIDIS, respectively, do not have a  
significant impact in our  estimates. This mainly happens because the  
most significant contributions to the effective $\chi^2$ minimized to  
obtain the replicas and to compute the new weights come, both for DIS and SIDIS,  from low-$x$ pseudo-data points where the effects of systematic uncertainties is diluted. In addition, for SIDIS at
higher values of $x$, the uncertainties are dominated by those of the  
fragmentation functions.
%
\subsection{Kinematical survey}\label{subsec:preview}

Our limited knowledge of the proton spin budget comes mostly from the rather limited kinematical coverage of the present spin dependent world data, the anticipated wide kinematical range projected for the EIC is of critical importance. It is very enlightening to start assessing the impact of the future EIC data studying the correlation and sensitivity coefficients \cite{Guffanti:2010yu,Wang:2018heo,Aschenauer:2019kzf} between the most representative helicity parton densities and the measured spin asymmetries, as a function of $x$ and $Q^{2}$. 

While the correlation coefficients indicate the regions of the phase space where an observable is expected to provide the strongest constraints on the partonic densities \cite{Guffanti:2010yu}, the ultimate impact on those distributions will also be determined by the present understanding of that observable. Therefore, it is instructive to analyze also the sensitivity coefficients \cite{Wang:2018heo} 
defined as a rescaled correlation that also accounts for the relation between the precision of the measurement and the current level of uncertainty of the parton density, which in this particular case is assumed to be given by the DSSV14 set. A more detailed discussion of the usefulness of these tools, as well as their applications to semi-inclusive data can be found in \cite{Aschenauer:2019kzf} and references therein. 

\begin{figure}[t]
\vspace*{0.1cm}
\begin{center}
\hspace*{-0.1cm}
\epsfig{figure=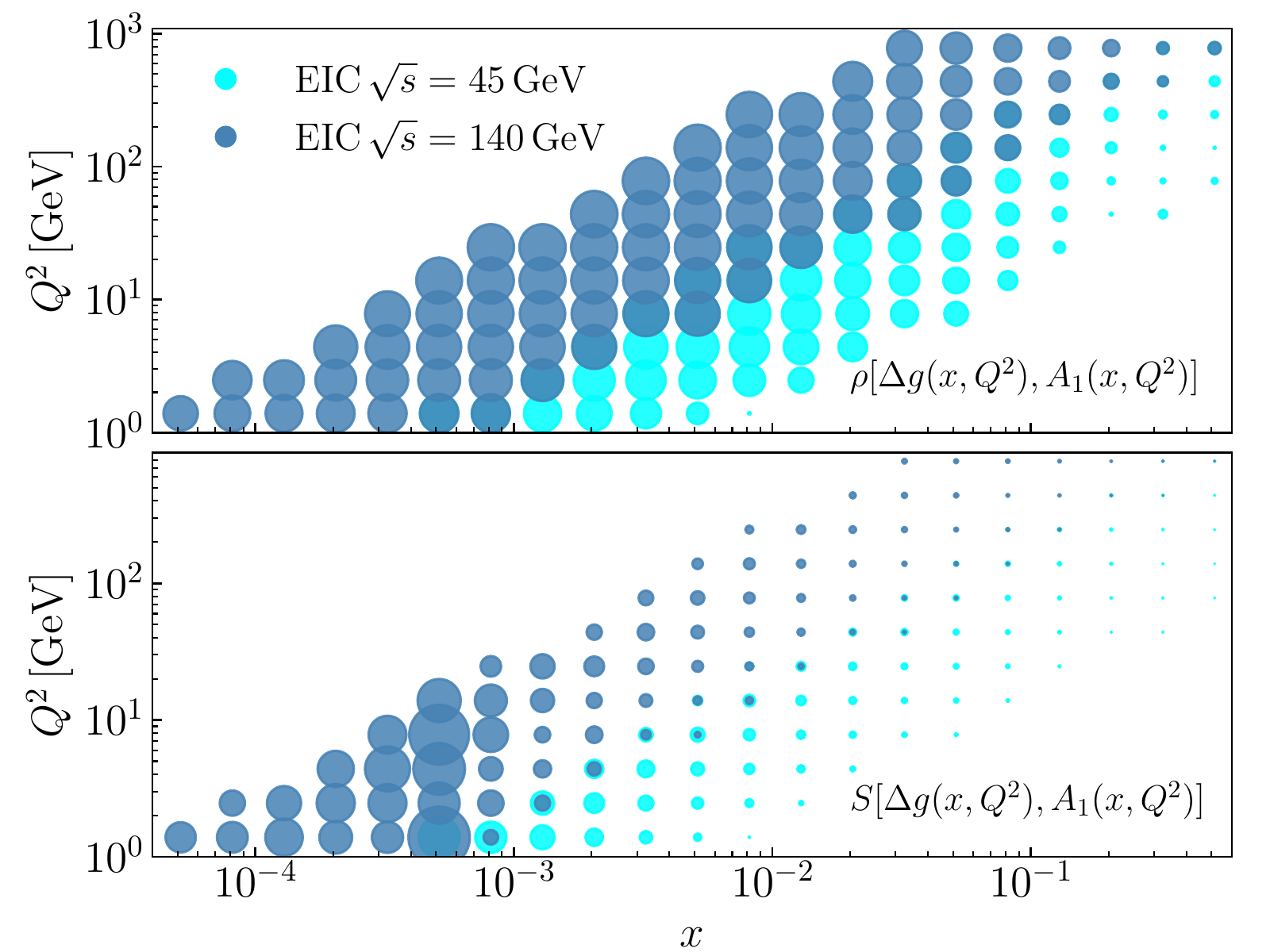,width=0.49\textwidth}
\end{center}
\vspace*{-0.5cm}
\caption{Correlation (upper panel) and Sensitivity (lower panel) coefficients between the gluon helicity distribution 
$\Delta g(x,Q^{2})$ and the double spin asymmetry $A_{1}$, as a function of $\{x,Q^{2}\}$. The light blue and blue circles 
represent the value of the correlation (sensitivity) coefficient for $\sqrt{s}=45$ GeV and $\sqrt{s}=140$ GeV, respectively. 
In all the cases the size of the circles is proportional to the value of the correlation (sensitivity) coefficient.}
\label{corr_map_gl}
\end{figure}

%

A major advantage of Monte Carlo sampling and using parton density replicas is the straightforward calculation of correlation and sensitivity coefficients between any projected measurement and a given parton distribution. 
The correlation coefficient $\rho\,[f_i ,{\cal O}]$ between a parton density for a given flavor $i$ and an observable ${\cal O}$ (i.e. a polarized asymmetry) is defined as \cite{Guffanti:2010yu}:
\begin{equation}
\label{eq:pdf_correlation}
\rho\,[f_i,{\cal O}]=\frac{\langle {\cal O}\cdot f_i\rangle -\langle {\cal O}\rangle \langle f_i\rangle }{\Delta{\cal O} \Delta{f_i}}\, ,
\end{equation}
where the mean values are calculated over the ensemble of replicas as  
\begin{equation}
\label{eq:mean}
\langle{\cal O} \rangle = \frac{1}{N} \sum_{k=1}^{N}{\cal O}[f_i^{(k)}]\,,
\end{equation}
with $N$ being the number of replicas, and the standard deviation for the observable and parton density  given by  
\begin{equation}
\label{eq:sigma}
\Delta {\cal O} = \sqrt{ \frac{1}{N-1} \sum_{k=1}^N \left({\cal O} [f_i^{(k)}] - \langle{\cal O} \rangle \right)^2  } 
\end{equation}
 Values for $|\rho|$ close to unity indicate that the observable and the parton density are highly 
correlated and therefore, the inclusion of these particular data with a competitive experimental 
uncertainty could be able to further constrain the parton density. Values close to zero are obtained 
for uncorrelated observables, and would never be able to improve the parton density determination,
irrespective of the precision  of the data. For simplicity, we omitted the dependencies on 
$x$, $Q^{2}$ and $z$, however the correlation coefficients are defined for the kinematics 
of each particular (pseudo)data point, allowing a straightforward comparison between the 
constraining power of different kinematics regions.
The sensitivity coefficient \cite{Wang:2018heo} on the other hand are given by
\begin{equation}
\label{eq:pdf_correlation_esc}
S[f_i,{\cal O}]=\frac{\langle {\cal O}\cdot f_i\rangle -\langle {\cal O}\rangle\langle f_i\rangle}{\xi\,\Delta{\cal O}\Delta{f_i}}\, ,
\end{equation}
where the scaling factor
\begin{equation}
 \xi \equiv \frac{\delta {\cal O}}{\Delta {{\cal O}}}
\end{equation}
is defined as the ratio of the experimental measurement uncertainty $\delta{\cal O}$, and the 
theoretical uncertainty for the same observable propagated from the parton density $\Delta{\cal O}$. 
The scaled correlation coefficient suppresses those regions of the phase space for which the 
experimental uncertainty is large compared to the uncertainty associated to the PDFs, 
while it enhances those regions where the relation is inverted, meaning where the 
biggest impact on the distributions is expected.

%
\begin{figure}[t]
\vspace*{0.1cm}
\begin{center}
\hspace*{-0.1cm}
\epsfig{figure=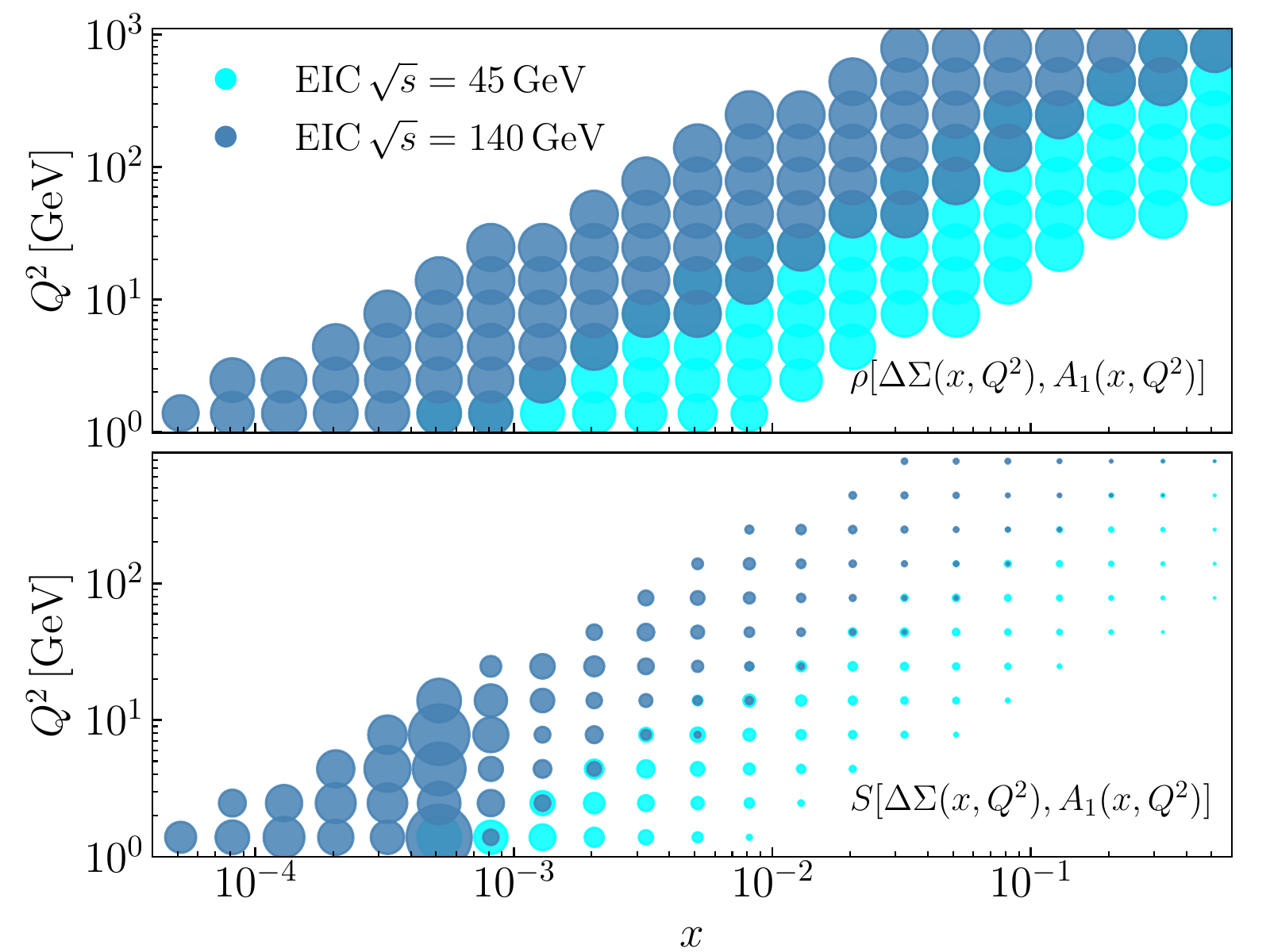,width=0.49\textwidth}
\end{center}
\vspace*{-0.5cm}
\caption{Same as Fig.~\ref{corr_map_gl}, for the quark flavor singlet helicity distribution, $\Delta\Sigma(x,Q^{2})$.}
\label{corr_map_sigma}
\end{figure}
Since inclusive photon mediated DIS data with proton beams by itself have no charge separation power, but
a limited flavor separation via the scale dependence, one expects to mainly determine the flavour singlet combination $\Delta \Sigma=\sum_{q=u,d,s}\,(\Delta q + \Delta \overline{q})$ and the gluon helicity distribution $\Delta g$. Consequently, we focus first on those helicity densities.   

In Figs.~\ref{corr_map_gl} and \ref{corr_map_sigma} we show the correlation (upper panels) and sensitivity (lower panels) coefficients between the gluon helicity distribution $\Delta g$ and the inclusive spin asymmetries $A_{1}$, as well as the correlation (sensitivity) between the flavour singlet helicity distribution $\Delta \Sigma$ and $A_{1}$,  as a function of $x$ and $Q^{2}$, for both of the c.m.s. energies under consideration, $\sqrt{s}=44.7 \,\mathrm{and}\, 141.4$ GeV. In all the cases, the size of the circles is determined by the value of the correlation (sensitivity) coefficient.

\begin{figure*}[h!]
\vspace*{0.1cm}
\begin{center}
\hspace*{-0.1cm}
\epsfig{figure=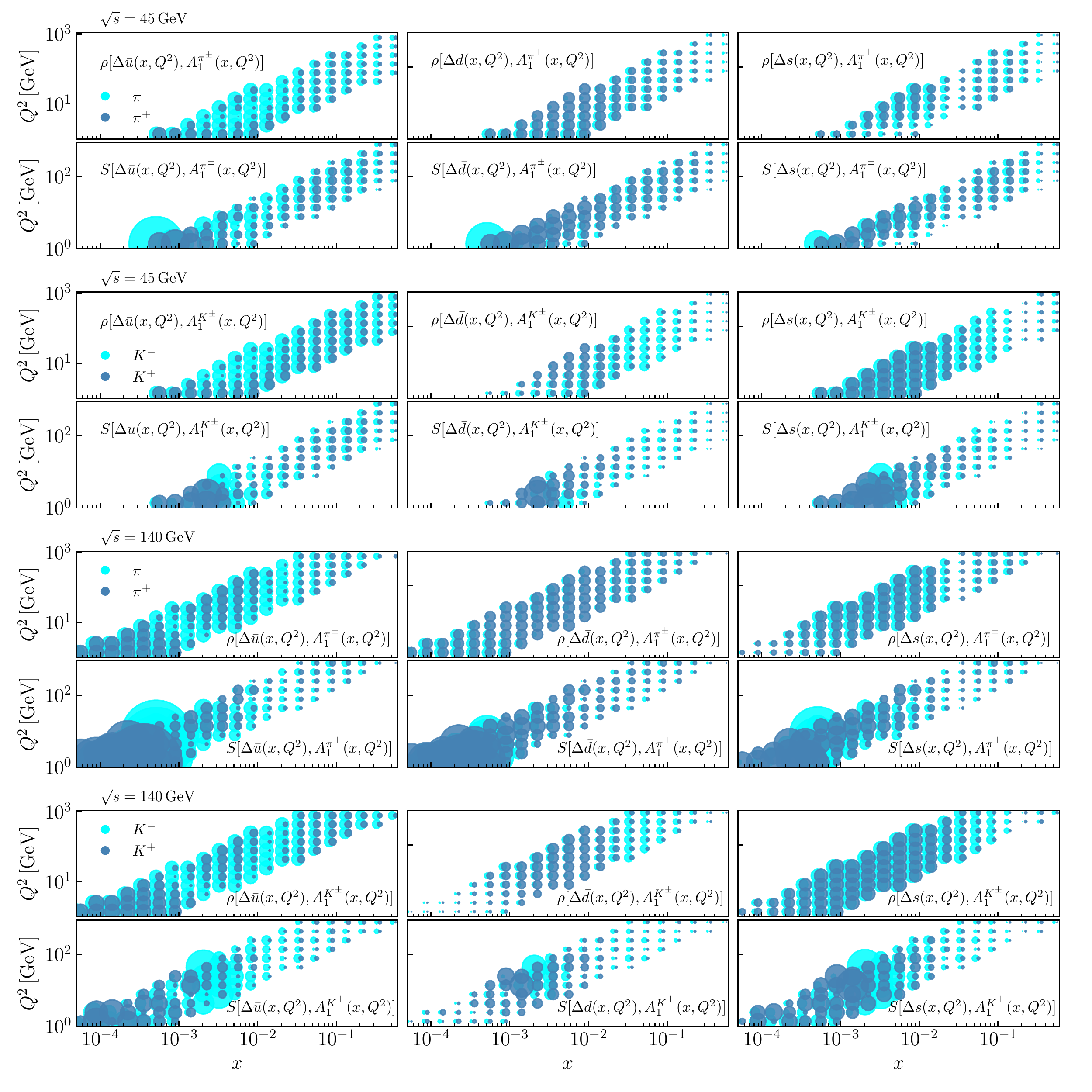,width=0.99\textwidth}
\end{center}
\vspace*{-0.5cm}
\caption{Correlation  and sensitivity coefficients between the sea quark helicity distributions 
and the pion and kaon SIDIS asymmetries as a function of $\{x,Q^{2}\}$. The light blue and blue circles represent the value of the correlation (sensitivity) coefficient for negatively and positively charged hadrons, respectively. In all the cases the size of the circles is proportional the value of the correlation (sensitivity) coefficient. The four upper rows correspond to $\sqrt{s}=44.7$ GeV SIDIS pseudo-data, 
and the lower rows to  $\sqrt{s}=141.4$ GeV, respectively}
\label{corr_SIDIS_pi}
\end{figure*}
The upper panel of Fig.~\ref{corr_map_gl}, shows that for both c.m.s. energy configurations there are strong correlations between the inclusive asymmetries and the gluon helicity, however the higher energy configuration covers one decade further in momentum fraction $x$. On the other hand, in the upper right corner of the plot, that corresponds to larger values of $x$ an $Q^2$, the correlation fades away faster for the lower energy configuration, reducing the grid of values over 
which the scale violations are effectively probed and suggesting a weaker impact. Focusing on the lower panel of Fig.~\ref{corr_map_gl}, the sensitivity plot shows that the lower left corner, mostly missed by the low energy configuration, is precisely where there is more potential for improving the gluon distribution. Notice that higher values of $x$ are already covered by RHIC measurements, and to lesser extent by fixed target DIS experiments.

For the singlet helicity distribution, $\Delta \Sigma$, the correlation and sensitivity coefficients show similar 
features however, at variance with what happens for the gluons, the correlation with the singlet remains strong in 
the upper right corner of the plot, due to the dominant quark contribution to the asymmetry at higher values of momentum fractions.
In this way we do not anticipate a significant difference in the constraint on $\Delta \Sigma$ coming from either 
energy configurations, except at very low values of $x$, only probed by the large c.m.s. energy configuration.

As we explain in more detail in Sec.~\ref{sec:res}, inclusive DIS asymmetries can not discriminate between quark and antiquarks, and therefore are unable to constrain by themselves the sea quark polarization, unless a very strong spin-flavor symmetry assumption is invoked. SIDIS asymmetries, on the other hand, weight the individual sea quarks contributions differently through their respective hadronization probabilities into specific final state charged hadrons. In Fig.~\ref{corr_SIDIS_pi} we show, precisely, the estimates for the correlation and sensitivity coefficients between the sea quark helicity distribution and the pion and kaon SIDIS spin asymmetries, reflecting these features.

Regarding the correlation plots for pion asymmetries, in the panels of the upper row in Fig.~\ref{corr_SIDIS_pi}, it is quite apparent that while $A_1^{\pi^-}$ is sensitive to $\overline{u}$ quarks, $A_1^{\pi^+}$ is sensitive to $\overline{d}$ quarks, being consistent with the valence quarks of the pion. The correlations are stronger at lower values of momentum fraction, where sea quark contributions 
dominate over those of the valence quark in the proton. Both pion asymmetries 
show a weaker but still significant correlation with strange quarks. The final impact of the measurements depends, of course, on a subtle balance between the uncertainties of the pseudo-data, the theory uncertainties inherent to the analysis, dominated in this case by the fragmentation functions, and our present knowledge on the sea quark polarization. This last features is quantified by the sensitivity coefficients shown in the second row panels of Fig.~\ref{corr_SIDIS_pi}, indeed suggest that a very 
significant impact should be expected. Notice that this estimates are based on our present knowledge on fragmentation functions. Since unpolarized SIDIS measurements at the EIC have been shown to improve significantly the determination of fragmentation functions \cite{Aschenauer:2019kzf}, the present estimates should be taken as rather conservative.

In the case of charged kaon SIDIS asymmetries, in the panels of the third row of Fig.~\ref{corr_SIDIS_pi}, show that the strongest correlations are found with strange quarks, to a lesser extent with $\overline{u}$ quarks and least with $\overline{d}$ quarks. 
This hierarchy can be again traced back to the parton composition of charged kaons, and to the implicit assumption in this analysis of the same polarization for strange quarks and anti-quarks. Finally, in terms of the sensitivity, the fourth row of panels of Fig.~\ref{corr_SIDIS_pi}, display that while the kaon SIDIS asymmetries have typically weaker impact on the $\overline{u}$ and $\overline{d}$ quarks than those with pions in final state, they have a much stronger sensitivity on the strange quark polarization.

The remaining rows in Fig.~\ref{corr_SIDIS_pi} show the same as the first four rows but for $\sqrt{s}=141.4$ GeV SIDIS data. The main difference with lower c.m.s. energy configuration is, as in the inclusive case, the data explores one decade lower in parton momentum fractions, where  the largest impact is expected.  This feature however, happens only for the quark flavors being a valence quark in the respective final state hadrons, as shown by both the correlation and sensitivity estimates, is is 
weakened for the unfavored quark types.

\section{Monte Carlo sampling of helicity parton distributions}\label{sec:MC}
%
In order to investigate the impact of the EIC data, as usual we implement a NLO global analysis combining the data sets coming from different spin dependent experiments performed so far with the pseudo-data discussed in the previous section. However, rather than using the Lagrange multiplier methodology to estimate uncertainties as in our previous assessments \cite{Aschenauer:2012ve,Aschenauer:2015ata}, in the following we implement a Monte Carlo sampling strategy with predetermined functional forms for the helicity densities, along the lines of reference \cite{deFlorian:2019zkl}, to which we refer for more specific details and discussions.  
The major advantage of this approach is to combine the computational power of the Mellin transform 
technique with the statistical tools provided by Monte Carlo sampling. Schematically, the strategy is to generate in a first step a Monte Carlo ensemble of replicas of the original data (and pseudo-data in the present case) with a probability distribution derived from the reported/expected uncertainties of each experiment within the desired accuracy \cite{DelDebbio:2007ee,Ball:2008by}. In a subsequent step, a set of parton densities is obtained for each replica of the data with the standard fitting techniques. The ensemble of corresponding parton densities replicas obtained in this way, is expected to encode all the information relevant to determine the parton distribution. For instance, the central value of the parton densities, 
or any quantity derived from them, is taken to be the average over the parton density replicas, and the corresponding uncertainty is the statistical standard deviation. In this way, the Monte Carlo sampling strategy avoids introducing a tolerance criterion to estimate uncertainties, and other known shortcomings in the propagation of uncertainties to experimental observables, characteristic to the Lagrange multiplier method \cite{Stump:2001gu} and the Hessian approach \cite{Pumplin:2001ct}, respectively.

\begin{figure}[t!]
\vspace*{0.1cm}
\begin{center}
\hspace*{-0.1cm}
\epsfig{figure=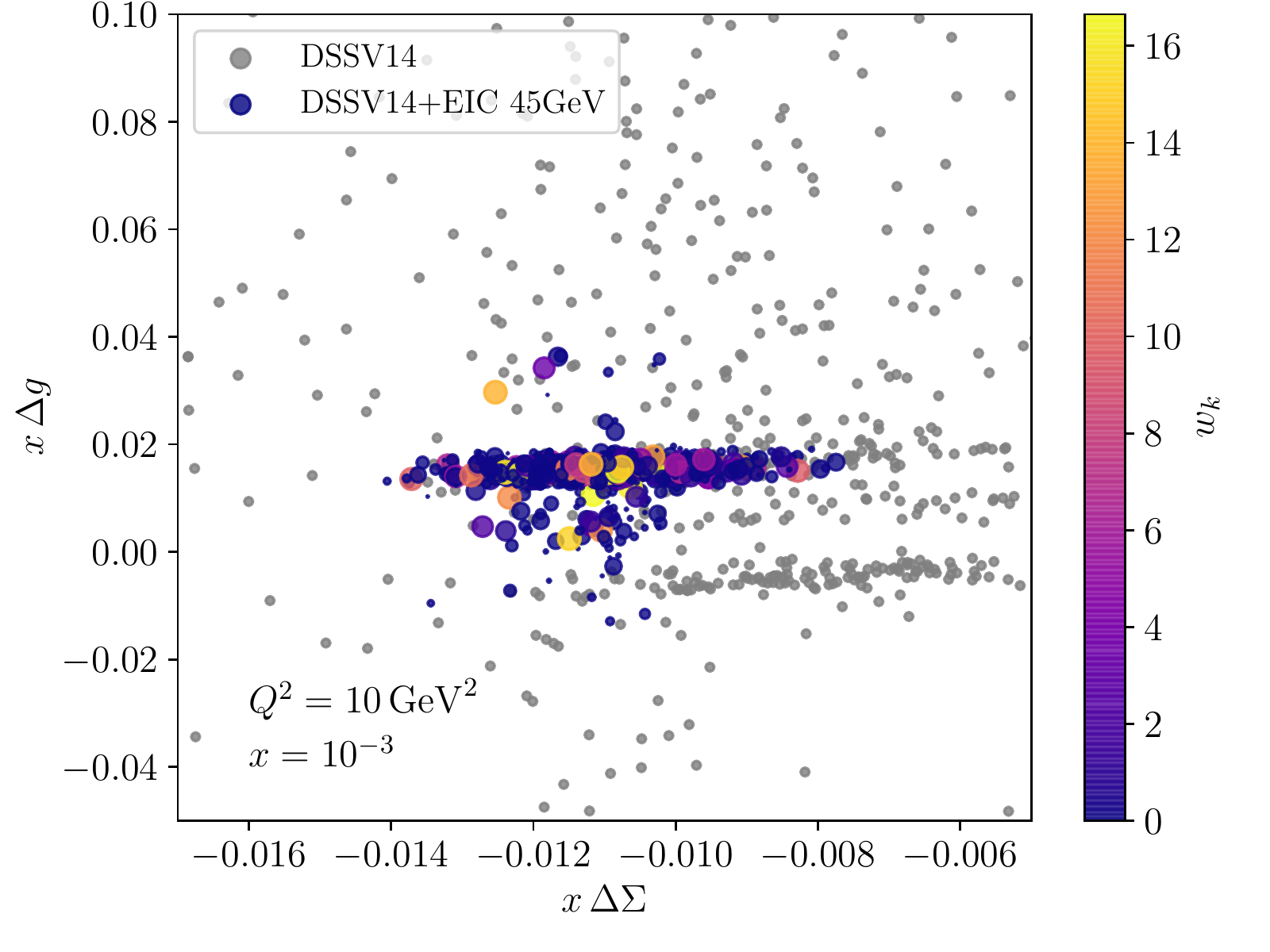,width=0.49\textwidth}
\end{center}
\vspace*{-0.5cm}
\caption{Values for $x\Delta g$ and $x\Delta \Sigma$  at $x=0.001$ and $Q^2=10$ GeV for the DSSV14 replicas (grey dots)and those of the new set that includes EIC inclusive DIS electron-proton  pseudodata at $\sqrt{s}=44.7$ GeV (in colour). The size and colour of the new replicas  represents the weight they obtain by reweighting with EIC $\sqrt{s}=141.4$ GeV pseudo-data. \label{fig:Gl_Q_scatter_d03}}
\end{figure}

The Monte Carlo sampling strategy provides also a very powerful tool to assess and compare the impact of different data sets in the determination of parton densities, known as reweighting. The reweighting technique \cite{Ball:2010gb,Ball:2011gg} allows to incorporate the information provided by a new set of data into an existing ensemble of parton density replicas without the need of refitting, but preserving 
the statistical rigor of its extraction. The method has already been successfully demonstrated in the context of the extraction of parton densities, see, for instance, Refs.~\cite{Nocera:2014gqa,Armesto:2013kqa,Paukkunen:2014zia,Borsa:2017vwy,deFlorian:2019zkl}.
By means of Bayesian inference, it is possible to modify the original probability distribution of an ensemble of parton density replicas to account for the information contained in a new measurement~\cite{Ball:2010gb}. This is implemented assigning a new weight to each replica, which measures its consistency with the new data. 

The Bayesian reweighting is equivalent to a refit including the additional set of data, as long as the impact of the new data is not too significant, for instance, by constraining some aspect of the parton densities that was largely undetermined before. Such a scenario would lead to a very large number of replicas with essentially vanishing weights making a full refit inevitable. This is precisely the situation of the EIC pseudo-data, since they extend by one or two additional decades in parton momentum fraction, respectively, as shown in the previous section. For this reason, rather than reweighting the DSSV14 replicas obtained in \cite{deFlorian:2019zkl}, we first produce a new set of replicas sampling the original DSSV14 data set supplemented with the inclusive DIS electron-proton  pseudodata at $\sqrt{s}=44.7$ GeV. 

In Fig.~\ref{fig:Gl_Q_scatter_d03} we present by the grey dots the results of the gluon helicity $\Delta g$ and the quark flavor singlet $\Delta \Sigma$ at $x=0.001$ at $Q^2=10$ GeV$^2$ as obtained from the DSSV14 replicas. The replicas are rather scattered over a comparatively large range of values of $\Delta g$ and $\Delta \Sigma$ that are representative of the corresponding uncertainties in both distributions. The coloured  dots represent the replicas obtained adding inclusive electron-proton DIS pseudo-data at $\sqrt{s}=44.7$ GeV.  The new replicas are clustered in a much more restricted area, which is the range of values that is the relevant one once the EIC data are included. Most of the original replicas become 
irrelevant, and those in the relevant region are extremely few in number.
Additionally, the size and colour of the new replicas represents the weight they obtain by reweighting with EIC inclusive DIS electron-proton pseudo-data at $\sqrt{s}=141.4$ GeV.

\begin{figure}[t!]
\vspace*{0.1cm}
\begin{center}
\hspace*{-0.1cm}
\epsfig{figure=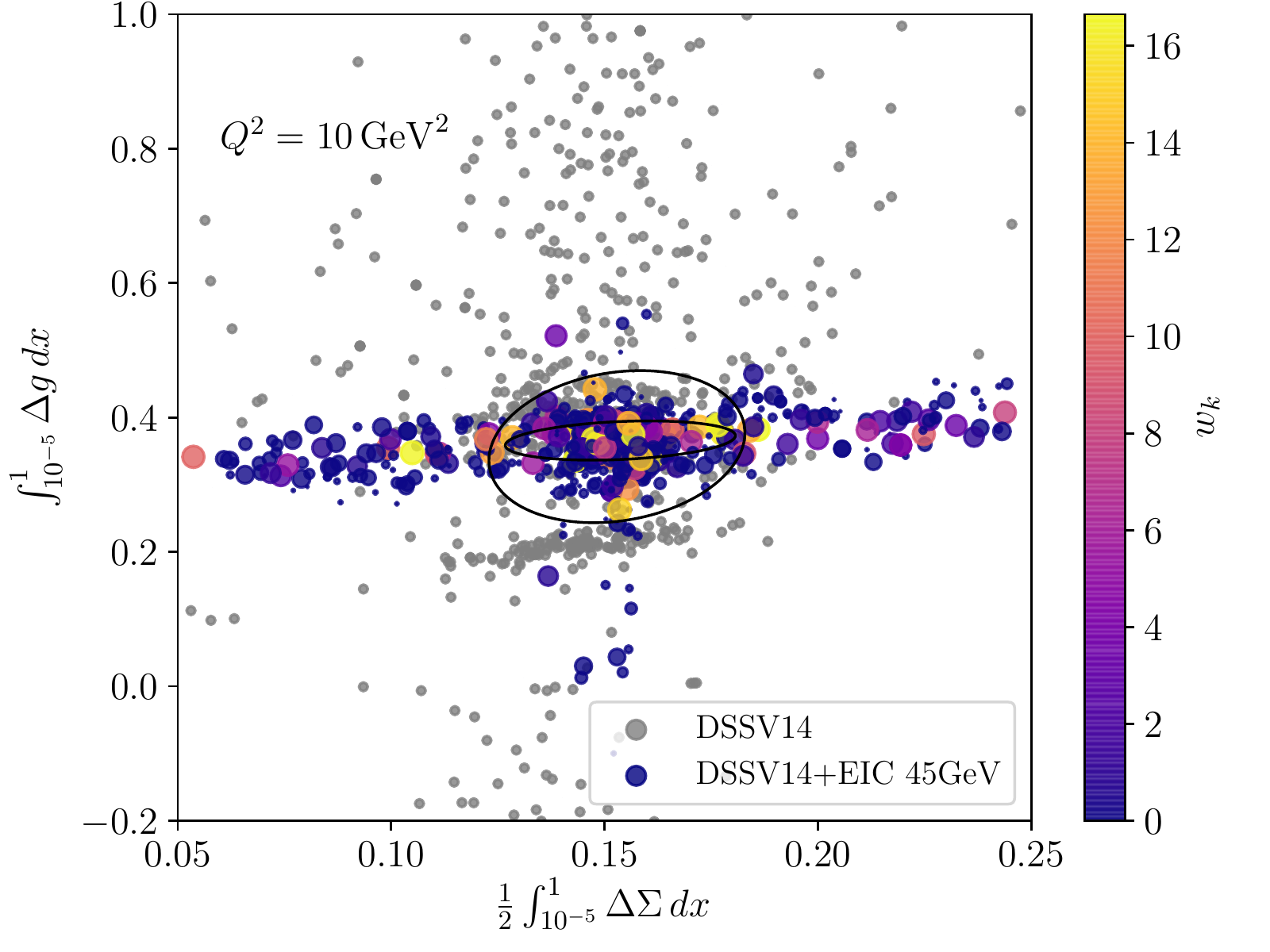,width=0.49\textwidth}
\end{center}
\vspace*{-0.5cm}
\caption{Values for the moments $\Delta g$ and $\Delta \Sigma$ integrated between $x=10^{-5}$ and $1$, at $Q^2=10$ GeV$^2$ for the DSSV14 replicas (grey dots) and those of the new set that includes EIC inclusive DIS electron-proton pseudodata at $\sqrt{s}=44.7$ GeV (in colour).
The size and colour of the new replicas is represents the weight obtained by reweighting with the EIC $\sqrt{s}=141.4$ GeV pseudo-data. \label{fig:Gl_Q_scatter_05}}
\end{figure}

In Fig.~\ref{fig:Gl_Q_scatter_05} show the same replicas as in Fig.~\ref{fig:Gl_Q_scatter_d03} but now in terms of the truncated moments $\Delta g$ and $\Delta \Sigma$, which represent the net contribution to the proton spin arising from the spin of quarks and gluons with momentum fractions larger than $x=10^{-5}$. These moments collect contributions not only from small $x$ but mostly from larger momentum fractions. The ellipses denote the 1-$\sigma$ limits for the values associated to the new replicas and their reweighting with DIS pseudo-data at $\sqrt{s}=141.4$ GeV 

Since we expect a much better constraining power coming from the EIC pseudodata at $x < 0.001$, as in Ref.\cite{deFlorian:2019zkl} we have further increased the flexibility of the parameterizations 
relative to that of the original DSSV14 set \cite{deFlorian:2014yva}, and enhanced the parameter sampling strategy to guarantee that no significant bias is introduced in the region constrained by the data, and that uniform probability distributions are obtained in the unmeasured region.

The DSSV analyses \cite{deFlorian:2008mr,deFlorian:2009vb,deFlorian:2014yva} adopt the most traditional fitting approach at NLO accuracy assuming a flexible functional form to parameterize the helicity PDFs as functions of the parton momentum fraction $x$ at an initial scale of $\mu_0=1\,\mathrm{GeV}$,   
\begin{equation}
x\Delta f_i(x,\mu_0) = N_i\, x^{\alpha_i} (1-x)^{\beta_i}
(1+\gamma_i \sqrt{x}+\eta_i x^{\kappa_i})\,,
\label{eq:param1} 
\end{equation}
where the label $i$ denotes different flavor combinations $\Delta u+\Delta \bar{u}$, 
$\Delta d+\Delta \bar{d}$, $\Delta \bar{u}$, $\Delta \bar{d}$, $\Delta \bar{s}\equiv\Delta s$, 
and the gluon density $\Delta g$. As usual, $\Delta f_i$ represents the difference of densities
with parton spins aligned and anti-aligned with the spin of the parent proton.
The optimization of the fit to data is carried out by varying the set of fit parameters 
$\{a_i\}=\{N_i,\alpha_i,\beta_i,\gamma_i,\eta_i, \kappa_i\}$ iteratively as long as a minimum in the effective $\chi^2$ function is reached. In each iteration the PDFs are evolved to the scale $\mu>\mu_0$ relevant in the experiment and used to compute the corresponding observables and the effective $\chi^2$ function to be minimized. 
Equivalently, Eq.(\ref{eq:param1}) can be rewritten to
\begin{equation}
x\Delta f_i(x,\mu_0) =\sum_{j=1}^3 N_{ij}\, x^{\alpha_{ij}} (1-x)^{\beta_{ij}}\,,
\label{eq:param2} 
\end{equation}
specially suited for working in the Mellin representation \cite{Stratmann:2001pb}, since each term is the integrand of an Euler integral of the first kind, and the corresponding moments are standard beta functions. Of course, some of the parameters in Eq.(\ref{eq:param2}) are no longer independent.  
This parameterizations have been found to be flexible enough to describe the DSSV14 data set in the 
sense that using more complex functional forms lead to equally good fits to data and also to statistically 
equivalent replicas of the data. Actually, the currently available data do not even fully constrain the values for the fit parameters and some restrictions on the parameter space have to be imposed, reducing them to typically five free paramenters per flavor or even less in the case of antiquarks, such that a unique and stable minimum in $\chi^2$ can be found. In spite of this flexibility in the region supported by the data, the values for the parameters that optimize the fit to data, determine the extrapolation into the unmeasured region, mostly $x<0.001$ and constrain artificially the range of variation of the distributions. To avoid this problem, additional terms in Eq.(\ref{eq:param2}) are added. The new parameters are chosen that they only modify the unmeasured low-$x$ domain, but leave the region constrained by the data unaffected, but it is required that the integrability of the parton densities and their convenient properties under mellin transformations are preserved. Different to Ref.\cite{deFlorian:2019zkl}, that   focused mainly on studying the already measured region $x>0.001$ and only one of these additional terms was included for the gluon helicity distribution, for the new replicas we allow three additional low-$x$ terms per flavor ($j=6$), allowing roughly a similar degree of flexibility in the so far unmeasured region $x<0.001$ as for the values of momentum fraction covered by the present data.  In this way 
we allow that even after the addition of the extremely precise EIC pseudo data at much lower values of $x$,
the replicas are allowed to vary in the new unmeasured region that is shifted about two decades in $x$.   


In the following section we present the results of the new fit, and the corresponding replica set, 
obtained combining the data set of the DSSV14 analysis with the inclusive DIS electron-proton pseudo-data  at $\sqrt{s}=44.7$ GeV.
In order to assess the impact of the remaining EIC pseudo data sets, such as the SIDIS measurements at $\sqrt{s}=44.7$ GeV, and inclusive DIS electron-proton and electron-helium at $\sqrt{s}=141.4$ and $\sqrt{s}=115.2$ GeV respectively, we then reweight the newly produced set of replicas. The outcome of these reweighting represents the combined impact of the first stage of the EIC together with the SIDIS measurements, with a second energy stage and including the results of electron-helium collisions, respectively. 

\section{Results}\label{sec:res}
\subsection{Impact of Deep Inelastic Scattering Data}\label{subsec:DIS}

\begin{figure}[b!]
\vspace*{0.1cm}
\begin{center}
\hspace*{-0.1cm}
\epsfig{figure=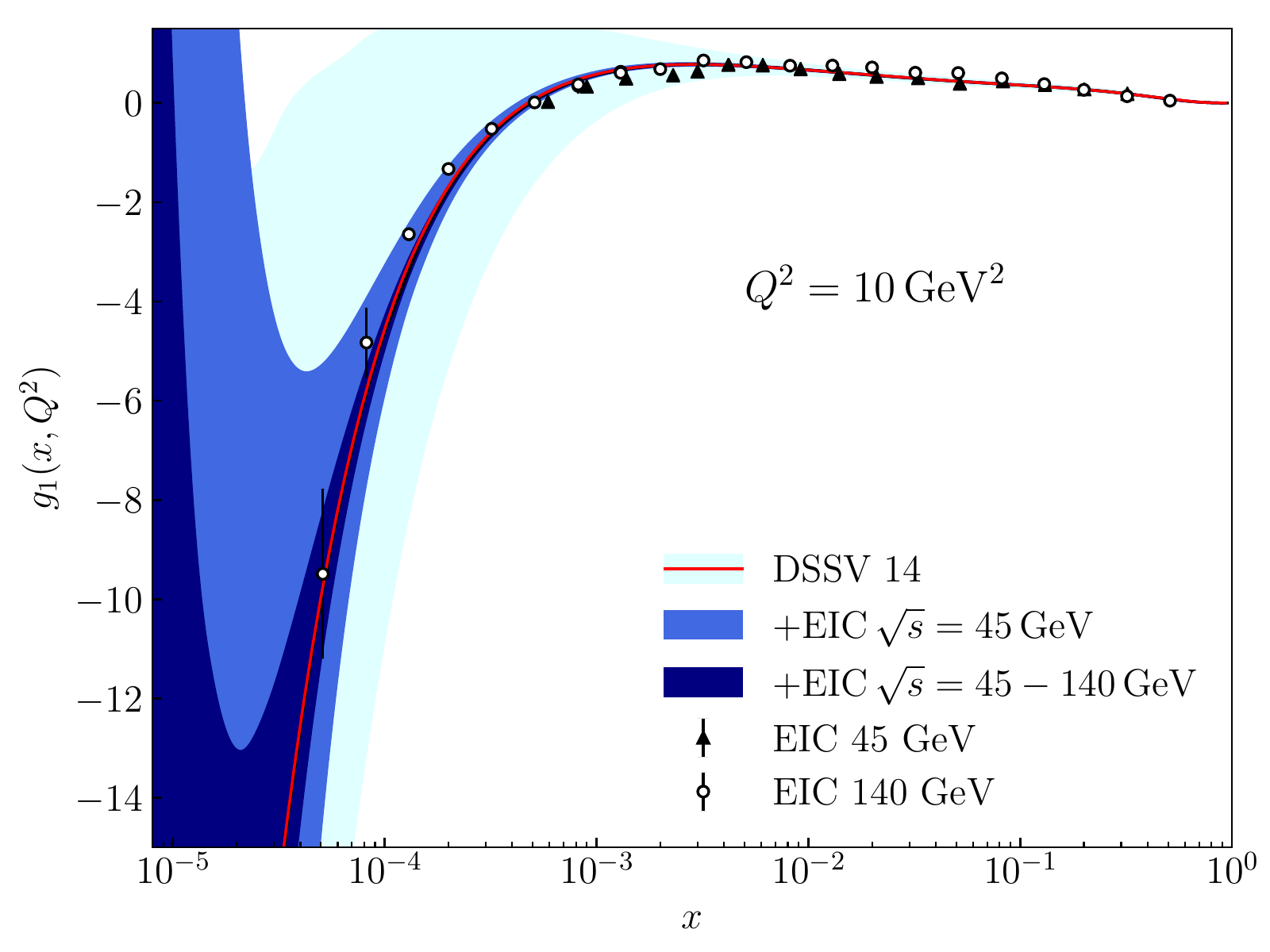,width=0.49\textwidth}
\end{center}
\vspace*{-0.5cm}
\caption{Helicity structure function $g_{1}(x,Q^{2})$ and its 68\% C.L. band as a function of $x$, at $Q^{2}=10$ GeV, calculated with the Monte Carlo variant of DSSV14. We include some the pseudo-data points of $g_{1}$ for the two c.m.s. energies and their expected experimental uncertainties. \label{fig:g1}}
\end{figure}

Our results focusing on the impact of the inclusive DIS measurements to the gluon helicity 
through the corresponding constraints on the spin dependent structure function $g_{1}(x,Q^{2})$ and its $Q^2$ dependence are described in the following. 
The totally inclusive photon mediated DIS cross section between longitudinally polarized leptons and nucleons can be written in terms of the structure function $g_{1}(x,Q^{2})$, in full analogy to the structure function $F_{1}(x,Q^{2})$, used in the case of the unpolarized cross section.

\begin{figure*}[t!]
\vspace*{0.1cm}
\begin{center}
\hspace*{-0.1cm}
\epsfig{figure=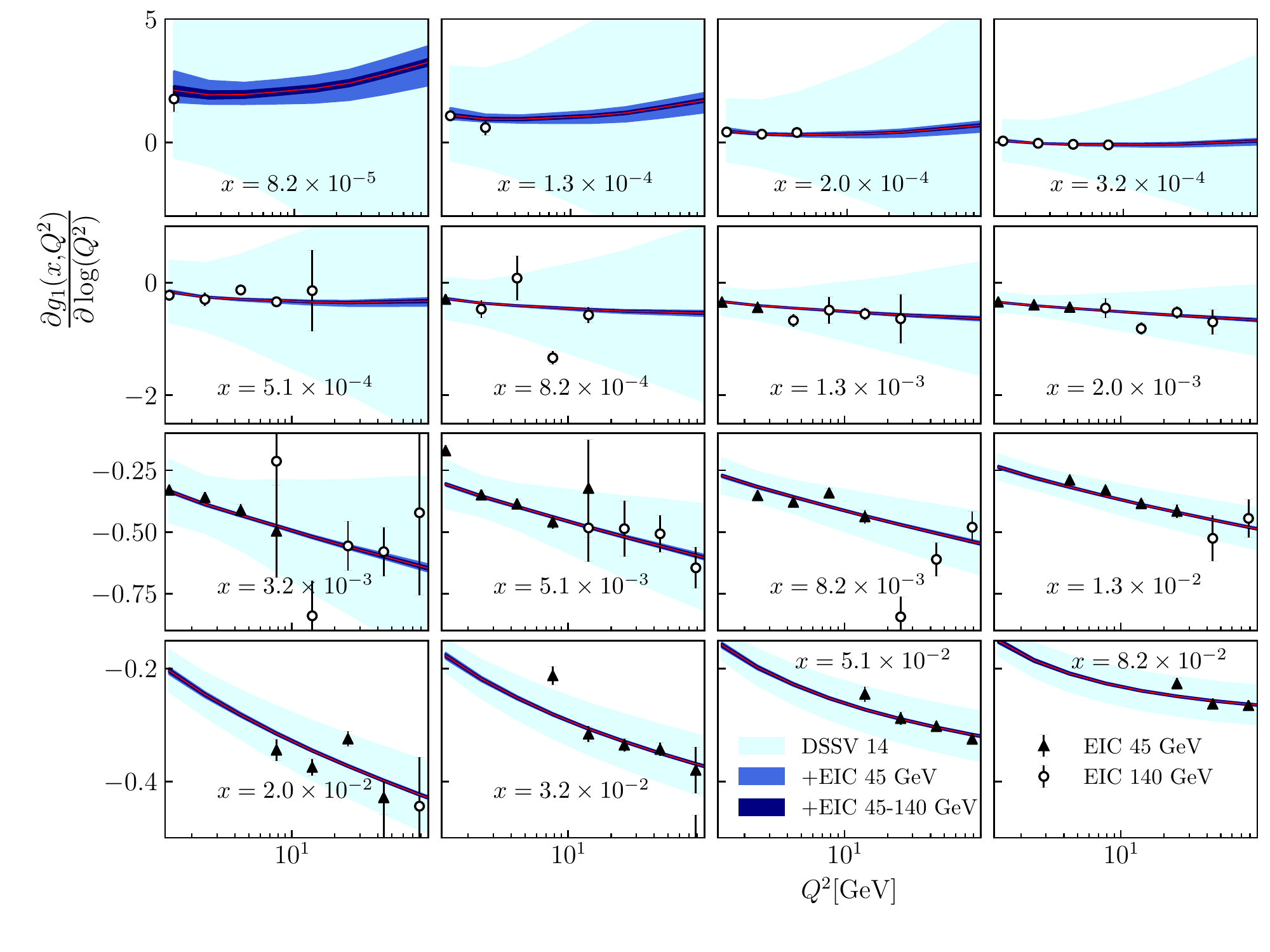,width=0.99\textwidth}
\end{center}
\vspace*{-0.5cm}
\caption{Estimates for the logarithmic scaling violation of $g_{1}(x,Q^{2})$ and the corresponding uncertainties, computed with the DSSV14 helicity parton densities, and the impact of including the DIS EIC pseudo-data sets at $\sqrt{s}=44.7$ GeV and $\sqrt{s}=141.4$ GeV, respectively
\label{fig:g1_derivlog}}
\end{figure*}

At next to leading order (NLO) in QCD, the structure function $g_{1}(x,Q^{2})$ is given in terms of the quark and gluon helicity densities $\Delta q(x,Q^{2})$ and $\Delta g(x,Q^{2})$ by,
\begin{eqnarray}
\label{eqg1} 
 g_{1}(x,Q^2)  =  \frac{1}{2} \sum_{q=u,d,s} e_q^2 \,
\left\{ \left(\Delta q + \Delta \overline{q} \right) \frac{}{} \right. \,\,\,\,\,\,\,\,\,\,\,\,\,\,\,\,\,\,\,\,\,\,\,\,\,\,\,\, \nonumber \\
+  \left. \frac{\alpha_s(Q^2)}{2\pi}   [ \Delta C_q \otimes
(\Delta q+ \Delta \overline{q} )  +  \Delta C_g \otimes
\Delta g ] \right\} ,
\end{eqnarray}
where $\Delta C_q$ and $\Delta C_g$ are the spin dependent DIS coefficient functions \cite{ap1}, $\alpha_s(Q^2)$ the QCD running coupling constant and $\otimes$ stands for the usual convolution integral. 
Restricting ourselves to three quark flavours, the structure function can be alternatively written as:
\begin{eqnarray}
\label{eqg1new}
 g_1x,Q^2)  = \left( \pm \frac{1}{12} \Delta q_3^{NS} + \frac{1}{36} \Delta q_8^{NS} 
+   \frac{1}{9} \Delta \Sigma \right) \nonumber \\ 
 \otimes     \left( 1 + \frac{\alpha_s}{2\pi}  \Delta C_q  \right)  
   +   \frac{\alpha_s}{2\pi} \sum_q e_q^2 \, \Delta C_g  \otimes  \Delta g,
\end{eqnarray}
where
\begin{eqnarray}
\label{eq38sigma}
\Delta q_3^{NS} &\equiv & \left( \Delta u +   \Delta \bar{u} \right) - \left( \Delta
  d +   \Delta \bar{d} \right) \nonumber \\
\Delta q_8^{NS} &\equiv& \left( \Delta u +   \Delta \bar{u} \right) + \left( \Delta
  d +   \Delta \bar{d} \right) -2 \left( \Delta s +   \Delta \bar{s} \right) \nonumber \\
\Delta \Sigma &\equiv& \left( \Delta u +   \Delta \bar{u} \right) + \left( \Delta
  d +   \Delta \bar{d} \right) + \left( \Delta s +   \Delta \bar{s} \right) 
\end{eqnarray}
and the $\pm$ sign in Eq.(\ref{eqg1new}) corresponds to scattering either a proton or a neutron, respectively. 
The first two non-singlet distributions, $\Delta q_3^{NS}$ and $\Delta q_8^{NS}$, evolve independently in $Q^2$ , whereas the gluon and the singlet distribution $\Delta \Sigma$ are coupled by the evolution equations.

It is worth noting that whereas $\Delta q_3^{NS}$ could be obtained directly from data on a linear combination of the proton and the neutron (helium, or deuterium) spin dependent structure functions, the remaining combinations in Eq.(\ref{eqg1new}) and $\Delta g$ have to be obtained indirectly, through their different scale dependences. 
The extended range both in $x$ and $Q^2$ of the data results that a  much more precise determination of all the distributions can be achieved. However, it should be kept in mind that even with an unbounded set of inclusive data of unlimited precision, it would be impossible to disentangle $\Delta q$ from $\Delta \bar{q}$. The latter requires data from processes that weight quark and antiquarks differently, such as weak interactions or processes including hadronizations, like those probed by SIDIS, where the 
flavor content of the hadron observed in the final state, can discriminate between quark and antiquark contributions.

\begin{figure}[b]
\vspace*{0.1cm}
\begin{center}
\hspace*{-0.1cm}
\epsfig{figure=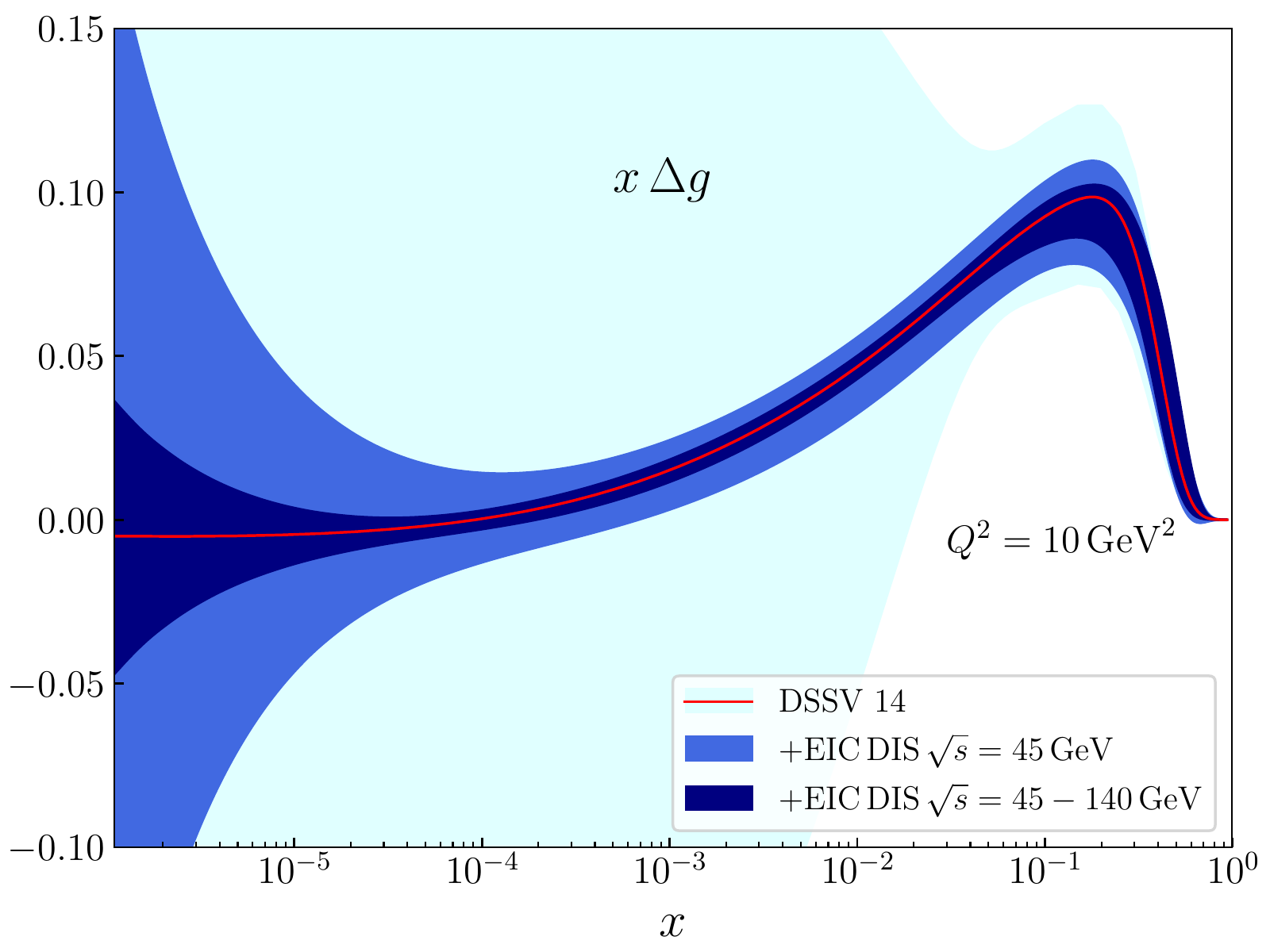,width=0.49\textwidth}
\end{center}
\vspace*{-0.5cm}
\caption{Impact of the projected EIC DIS pseudodata on the gluon helicity distribution. Together with the DSSV14 estimate, we show the uncertainty bands resulting from the fit that includes the $\sqrt{s}=44.7$ GeV DIS pseudo-data, and the reweighting with $\sqrt{s}=141.4$ GeV pseudo-data \label{gl_refit}}
\end{figure}
%
In Fig.~\ref{fig:g1} we show the structure function $g_{1}(x,Q^{2})$ as a function of $x$ and for a fixed value of $Q^{2}=10$ GeV, computed at NLO accuracy using the Monte-Carlo variant of the DSSV14 set of helicity distributions \cite{deFlorian:2019zkl}, which may be taken as representative of our present knowledge. The light cyan band represents its uncertainty for the 68\% C.L. limit. The world data on $g_{1}$, is actually restricted to $x>0.004$ and consequently below that threshold the DSSV14 expectation is just an extrapolation assuming that $\Delta q$ and $\Delta g$ vanish smoothly as $x \rightarrow 0$. 
The lack of data is reflected in the very rapid growth of the uncertainty band. 
Overlaid with the expectation for $g_{1}$, we also show some representative EIC pseudo-data points, generated for both of values of the c.m.s. energy analyzed in this study. The points are plotted 
just as a reference for the corresponding ranges in $x$, and for the size of the expected uncertainties. Notice that the pseudo-data points occur at different values of $Q^{2}$, the selected points 
shown in the plot are those closest to $Q^{2}=10$ GeV.

The light blue band in Fig.~\ref{fig:g1} shows the uncertainty in the structure function estimate when the 
inclusive DIS electron-proton pseudodata at $\sqrt{s}=44.7$ GeV are combined with the DSSV14 data set, and represent the comparative impact of the projected measurement. The darker blue band shows the effect of reweighting the replicas with DIS inclusive electron-proton pseudo-data at $\sqrt{s}=141.4$ GeV, and accounts for the combined effect of both pseudo-data sets. The reweighting yields a number of effective replicas $N_{eff}=82$ to be compared with a total of $950$, which means a very significant impact.

\begin{figure}[b!]
\vspace*{0.1cm}
\begin{center}
\hspace*{-0.1cm}
\epsfig{figure=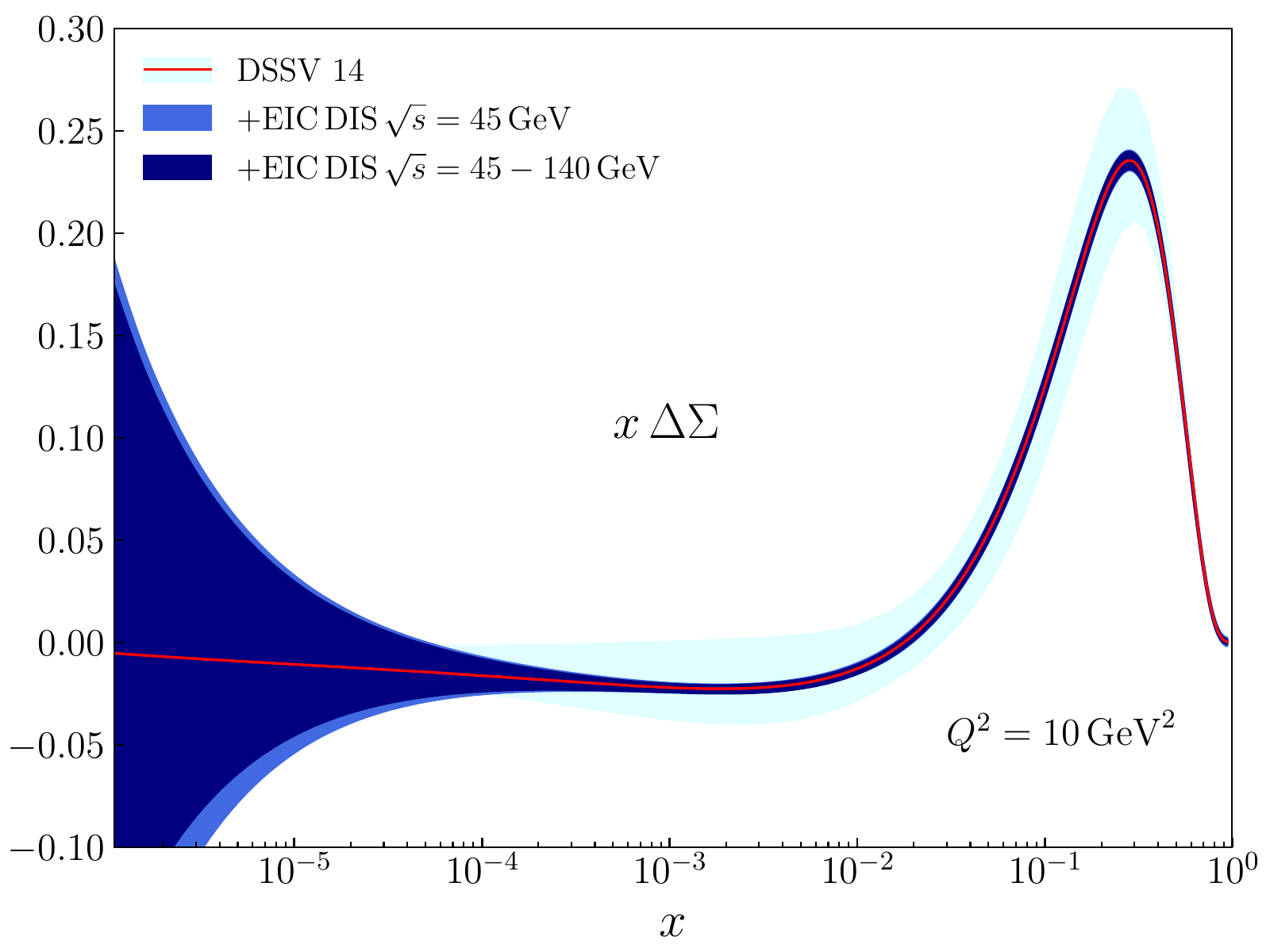,width=0.49\textwidth}
\end{center}
\vspace*{-0.5cm}
\caption{The same as Fig.~\ref{gl_refit} but for the singlet quark helicity $\Delta \Sigma$ and its uncertainty bands at $Q^2=10$ GeV$^2$ 
\vspace{11mm} \label{sigma_refit}}
\end{figure}

Even though the exact scale dependence of $g_{1}$ is related to the size and evolution the whole set of helicity distributions in an extremely convoluted way, at small enough values of $x$, it is linked to the gluon helicity through the approximate relation:
\begin{equation}
\frac{\partial g_{1}(x,Q^{2})}{\partial \ln Q^{2}}\approx -\Delta g(x,Q^{2}),
\end{equation}
it simply states for a given value of $x$ the larger is the gluon helicity, the steeper is the $Q^2$-dependence of $g_{1}(x,Q^{2})$, and it gives an more intuitive picture of how an improved knowledge of $g_{1}$ at different scales, constrains the gluon helicity.

In Fig.~\ref{fig:g1_derivlog} we present the estimate for the logarithmic derivative of $g_{1}$ and the corresponding 68\% C.L. limit uncertainty bands, as a function of $Q^{2}$ for different values of 
$x$. The values for the logarithmic derivative of $g_{1}$ in Fig.~\ref{fig:g1_derivlog} were estimated approximating the derivative as a ratio between the differences of two consecutive pseudo-data points, and the logarithms of their corresponding $Q^2$ values. Therefore, for each bin in $x$ it is necessary to have at least to pseudo-data values for $g_{1}$ in different bins of $Q^{2}$. These approximations are used just to visualize the correlation. In the analysis we use always the full NLO evolution.
Similarly to Fig.~\ref{fig:g1}, we present the expectations derived from EIC pseudo-data for the two values of $\sqrt{s}$, as well as the impact (in terms of the uncertainty bands) that those data points would have through the new replicas and their reweighting.  

\begin{figure*}[t!]
\vspace*{0.05cm}
\begin{center}
\hspace*{-0.1cm}
\epsfig{figure=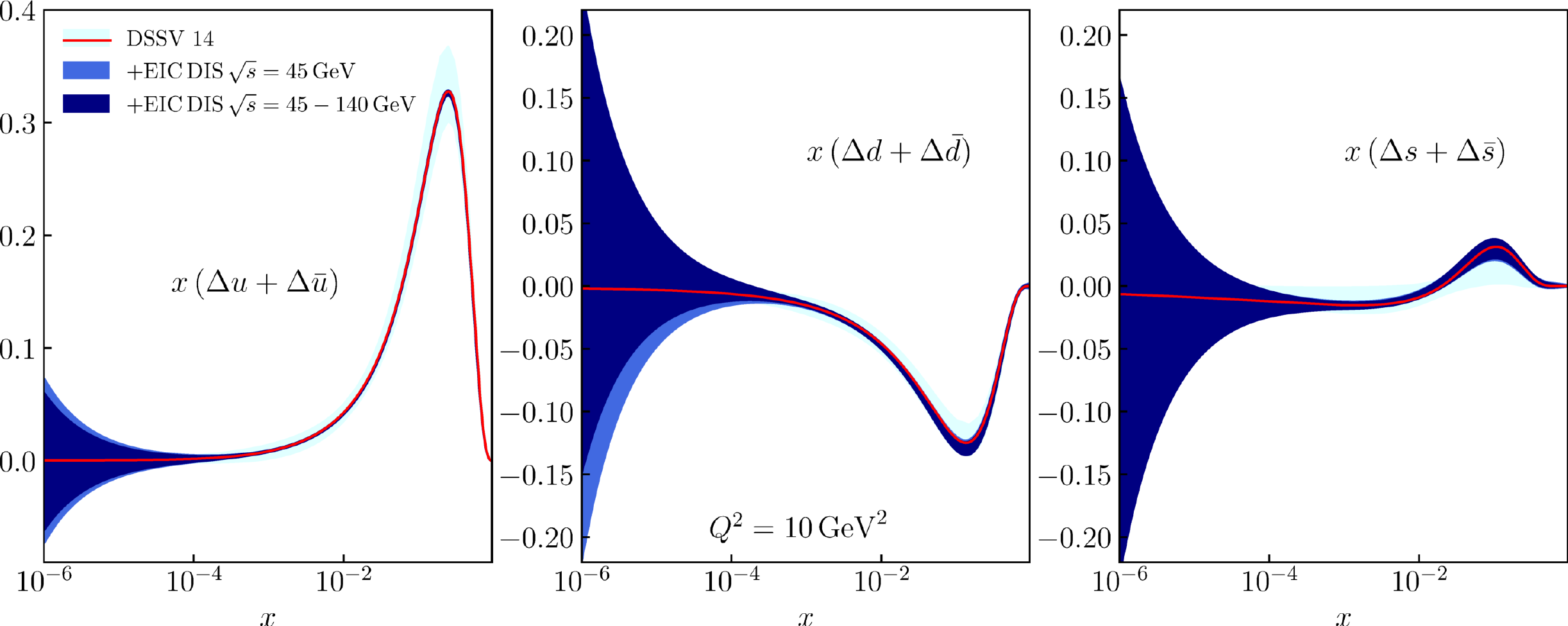,width=0.99\textwidth}
\end{center}
\vspace*{-0.5cm}
\caption{Impact of the projected EIC DIS data on the the total quark helicity distributions $\Delta u + \Delta \overline{d}$, $\Delta d + \Delta \overline{d}$ and $\Delta s + \Delta \overline{s}$, respectively. Together with the DSSV14 estimate, we show the uncertainty bands resulting from the fit that includes the $\sqrt{s}=44.7$ GeV DIS pseudo-data, and the reweighting with data at $\sqrt{s}=141.4$ GeV.  \label{dist_dis_refit}}
\end{figure*}

Let us make some remarks at this point. In the first place, and as expected, the uncertainty in the 
scaling violation grows dramatically in the DSSV14 estimate for lower values of $x$, due to the lack of data and therefore constraints to the gluon helicity for $x\lesssim 10^{-2}$. Second, the EIC pseudodata reduce considerably the range of variation allowed in the slope of $g_{1}$, and consequently the value for $\Delta g$. Finally, the difference in the $x$-range covered by the data for different c.m.s. energies is significantly different and therefore critical, since the most important constraints on the gluon distribution are expected to come from the the region where the scaling violations are measured, as will be discussed below.

In terms of the helicity gluon distribution itself, the impact of the projected EIC data  is even more graphic. In Fig.~\ref{gl_refit} we show the gluon helicity distribution and its uncertainty bands, as a function of $x$, for $Q^{2}=10$ GeV. The uncertainty estimates correspond to the standard deviation of the DSSV14 replicas (in light cyan), that of the replicas obtained combining the DSSV14 data set and the $\sqrt{s}=44.7$ GeV pseudo-data set (royal blue), and after reweighting the latter with the $\sqrt{s}=141.4$ GeV set (dark blue). As anticipated, the impact on the gluon helicity uncertainty is very significant, not only at very low values of momentum fraction where the DSSV14 is basically 
unconstrained but also at intermediate values, where in spite of the availability of inclusive DIS data in the DSSV14 data, the gluons are still largely unconstrained. The impact of the combined high and low c.m.s. energy configurations is a reduction in the uncertainty roughly by a factor between three and four and pushes the growth of the uncertainty bands one decade lower in $x$ , that eventually happens when the pseudodata become scarce or have less span in $Q^2$.

Similar considerations are in order regarding the impact of DIS pseudodata on the singlet quark helicity distribution $\Delta \Sigma$, which is shown in Fig.~\ref{sigma_refit}. In this case, however, the distribution is already much better constrained by fixed-target DIS experiments integrated in the DSSV14 analysis. Nevertheless, the impact of the much more precise EIC pseudodata is very significant. Different to what happens for the gluon helicity, the addition of the high c.m.s. energy pseudo data set is not as significant, except at very low values ofthe momentum fraction. 

\begin{figure*}[t!]
\vspace*{0.05cm}
\begin{center}
\hspace*{-0.1cm}
\epsfig{figure=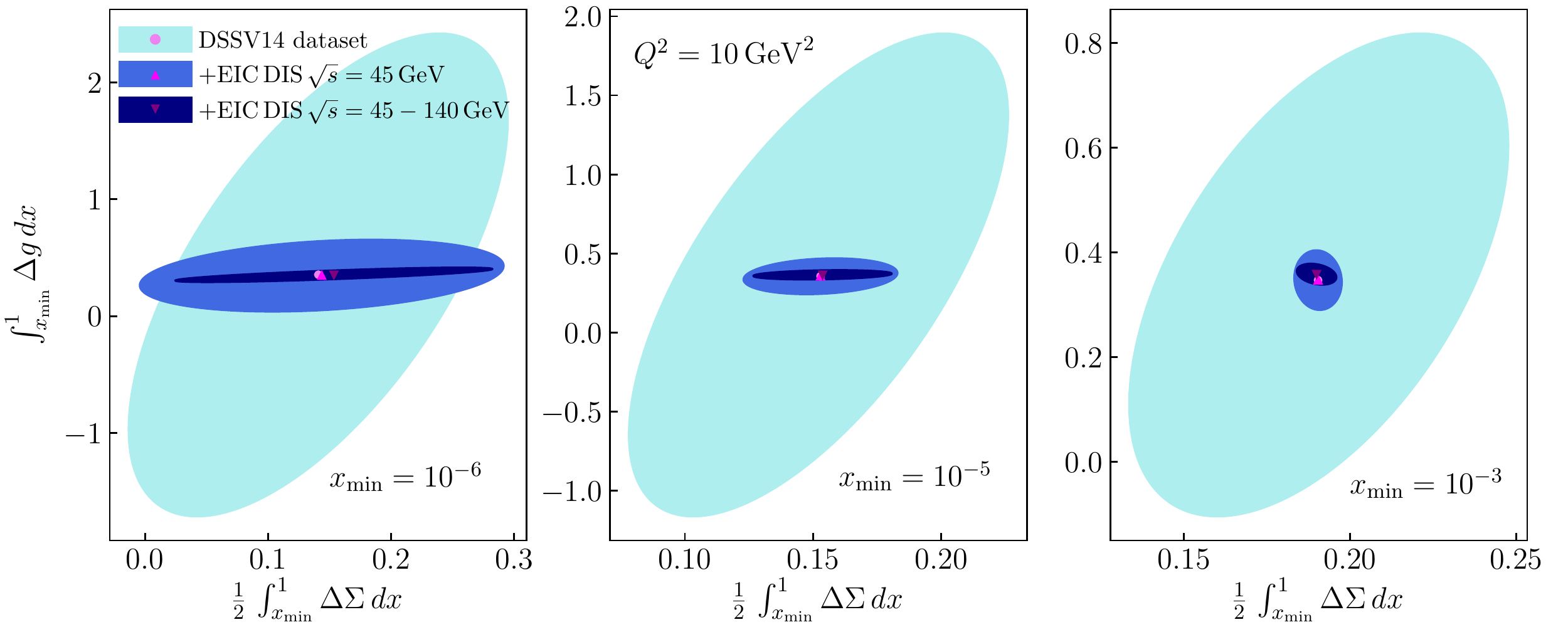,width=0.99\textwidth}
\end{center}
\vspace*{-0.5cm}
\caption{Correlated uncertainties for the truncated integrals of $\Delta g$ and $\Delta \Sigma$ in the range $10^{-6}<x<1$ (left panel), $10^{-5}<x<1$ (center panel) and $10^{-3}<x<1$ (right panel) computed with DSSV14 (outer area), as well as including pseudo-data from the EIC (inner areas). \label{papa_trifecta}}
\end{figure*}

For completeness in Fig.~\ref{dist_dis_refit} we show the helicity distributions  $\Delta u + \Delta \overline{u}$, $\Delta d + \Delta \overline{d}$ and $\Delta s + \Delta \overline{s}$. As in the case of $\Delta \Sigma$, there is a significant improvement compared to the DSSV14 analysis in region of $x$ where both data and pseudo-data overlap ($x>0.001$). This region combines the precision of EIC pseudodata with the flavor discriminating power of the original DSSV14 data set. 
Again, the difference between the two c.m.s. energy configurations is not as relevant as in the case of the gluon helicity, because the flavor separation has already been achieved, except at very low $x$.
As usual, the $\Delta u + \Delta \overline{u}$ distribution is much better constrained than those for the other flavors. This feature can be simply traced back to the electric charge factor for $u$ quarks in the proton structure function $g_{1}$, which is four times larger than those for $d$ and $s$ quarks. On the other hand, the latter show rather similar uncertainties at very small-$x$. It is noted that both in DSSV14 and in the variants enhanced with EIC pseudodata we assume $\Delta s = \Delta \overline{s}$.  

Rather than the total quark helicities, the quark flavor singlet combination $\Delta \Sigma$, or the gluon helicity distribution $\Delta g$, and their the moments, i.e. their integrals over the parton momentum fraction, have a much more simple and intuitive interpretation: the net contribution of quarks and gluons to the proton spin. It is very instructive then to check how well determined are these quantities, and how their values and corresponding uncertainty estimates depend on the range of momentum fraction evaluated. Both quantities are typically correlated so it is natural to represent them in a two dimensional plot, as shown in Fig.~\ref{papa_trifecta}.

Going from right to left in Fig.~\ref{papa_trifecta}, the rightmost panel shows the correlated uncertainty of the moments of the gluon (vertical axis) and the quark singlet helicity distributions (horizontal axis) integrated down to $x_{min}=0.001$, or in other words, the net contribution to the proton spin coming from gluons and quarks down to one thousand of the proton momentum. The light cyan area represents the range  allowed by the data included in the DSSV14 analysis, while the light and darker blue areas denote the impact of the projected EIC measurements. Notice once again, that the main difference between the impact of both c.m.s. energy configurations is in the constraint on the gluon contribution to the proton spin, roughly by a factor of two.

Clearly, quarks and gluons with smaller momentum fractions may also contribute to the proton spin, and since parton densities tend to be increasingly less constrained towards smaller $x$, the uncertainties on the moments typically grow, as shown in the center and leftmost panels, where the integrals are carried  down to $x_{min}=10^{-5}$ and $x_{min}=10^{-6}$, respectively. 
EIC pseudo-data stop slightly above $x=10^{-5}$, and below that threshold both the distributions and the moments should be only marginally limited by the continuity, integrability, and positivity of the helicity distributions. 

In the DSSV14 low-$x$ extrapolation scenario, the integrals of the singlet quark and gluon helicity distributions saturate rather early, consistent with a picture where at very low-$x$ partons become unpolarized. Therefore the central values of the truncated moments do not change significantly as lower values are considered for the lower integration limit $x_{min}$. 
This feature may or may not be validated by the future EIC measurements, and the chance to verify such behaviour emphasizes the relevance of scanning the progression of the moments down to the lowest possible values with the smallest uncertainties. 

\begin{figure}[b!]
\vspace*{0.05cm}
\begin{center}
\hspace*{-0.1cm}
\epsfig{figure=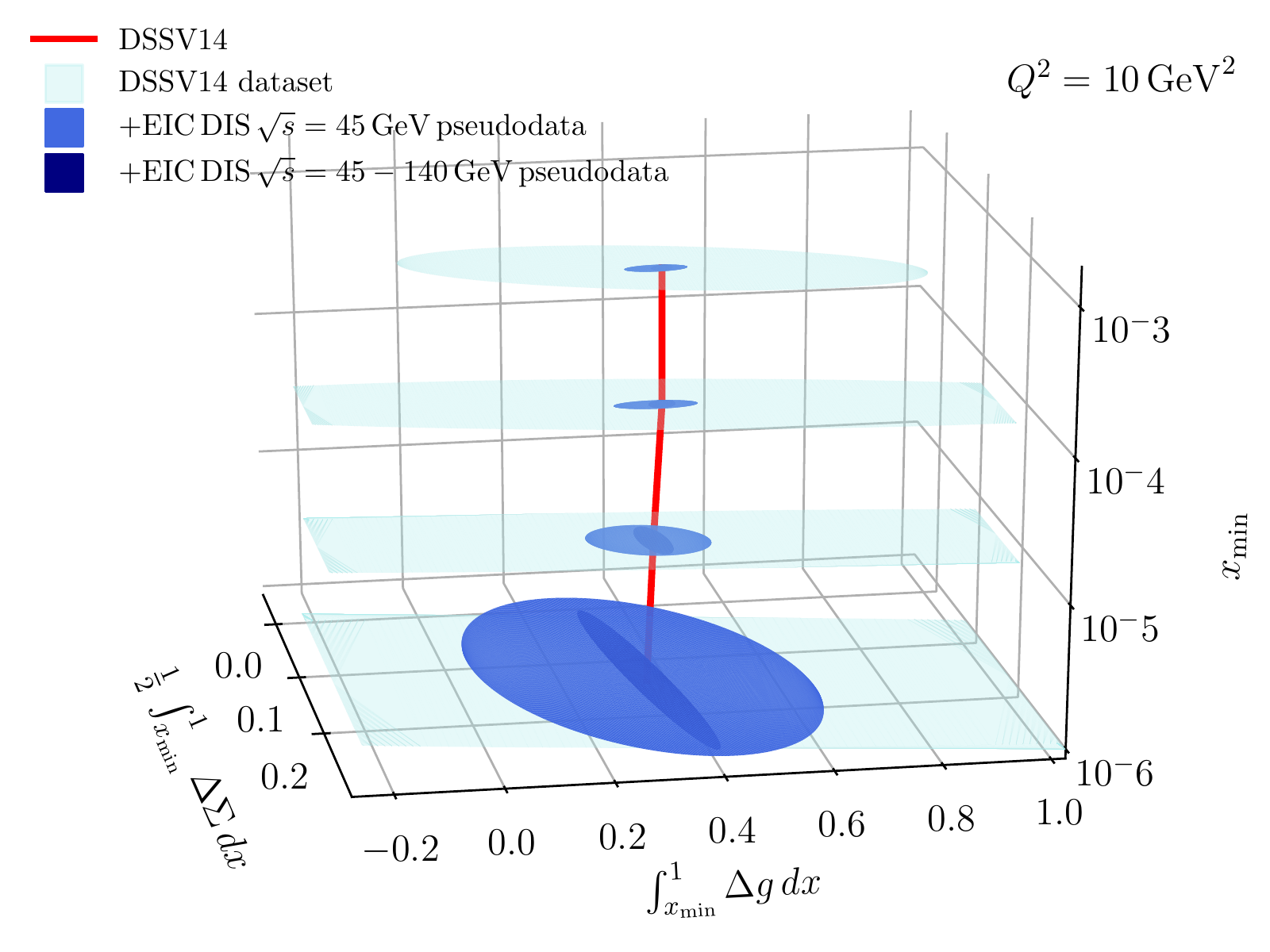,width=0.49\textwidth}
\end{center}
\vspace*{-0.5cm}
\caption{The same as Fig.\ref{papa_trifecta}, but in a common scale for comparison. The red curve represents the central values of the truncated moments as a function of $x_{min}$, computed extrapolating the DSSV14 scenario. 
\label{papa_slices}}
\end{figure}

In Fig.~\ref{papa_slices}, we show again the three area plots of Fig.~\ref{papa_trifecta} but now in the same scale, together with the correlated uncertainty of the moments truncated at $x_{min}=10^{-4}.$  The central value of the truncated moments as a function of $x_{min}$ computed in the DSSV14 extrapolation scenario is plotted as a red curve. A sizable change in the slope of the red curve would suggest that wee partons play a significant role in the proton spin.

\begin{figure}[t]
\vspace*{0.05cm}
\begin{center}
\hspace*{-0.1cm}
\epsfig{figure=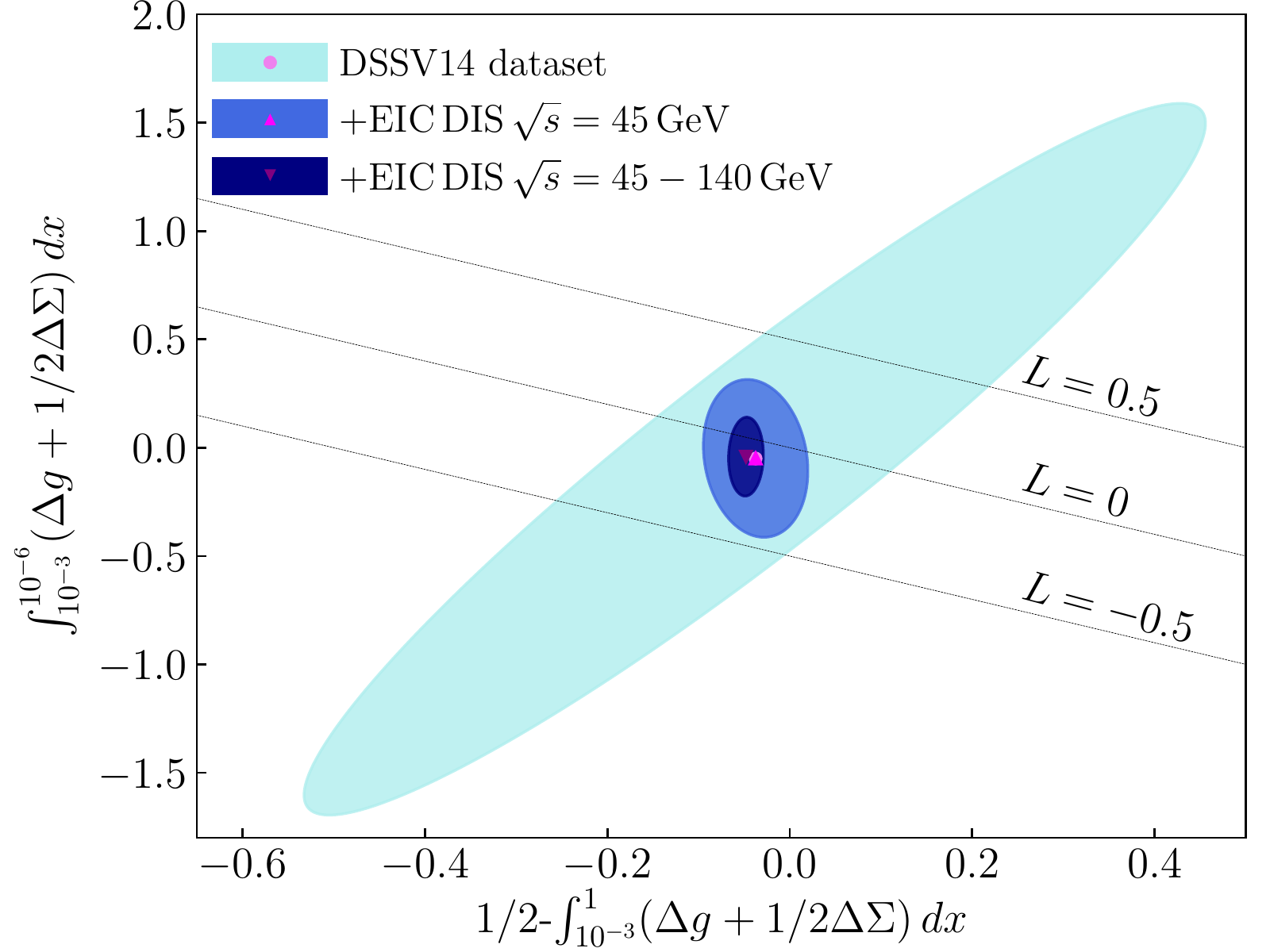,width=0.49\textwidth}
\end{center}
\vspace*{-0.5cm}
\caption{Room left for potential orbital angular momentum contributions to the proton spin according to present data and future EIC measurements.\label{missing_L}}
\end{figure}

Once we have a precise estimate for the constraints that EIC will impose on the net quark and gluon contribution to the proton spin, it is natural to ask how much room they will leave for other contributions, like the one that could come from quark and gluon orbital angular momentum. For instance, one could assume that the difference between the net quark and gluon spin contribution and the actual proton spin 1/2, could precisely be the contribution from the orbital angular momentum. 
This is represented in Fig.\ref{missing_L}. In the horizontal axis we show the difference between 1/2 and the contribution from the spin of quarks and gluons for a momentum fraction down to $x=0.001$. This would be the room left to the orbital angular momentum if the net spin contribution from partons with smaller momentum fractions is very small or even zero. 
But as the latter could actually be non negligible, and is currently very uncertain, we represent in the vertical axis their potential contribution. The colored areas show the constraints on these values coming from present data, in light cyan, and those that one would expect from the projected EIC measurements. The diagonal lines represent the combinations of low and high $x$ contributions for which the resulting orbital angular momentum would be as large as the proton spin and parallel to it, vanishing, or exactly opposite. The EIC data would be able to discard at least one of these extreme scenarios, and perhaps two of them.

\begin{figure*}[t!]
\vspace*{0.05cm}
\begin{center}
\hspace*{-0.1cm}
\epsfig{figure=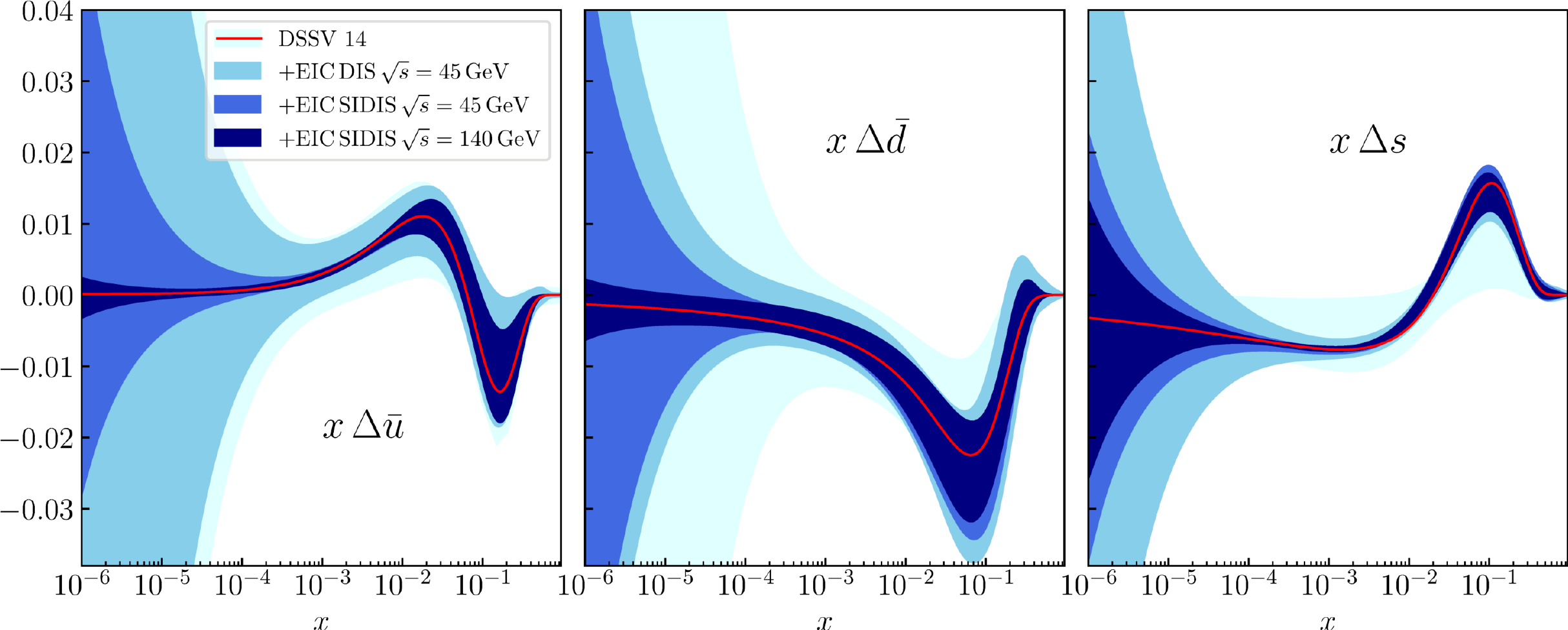,width=0.99\textwidth}
\end{center}
\vspace*{-0.5cm}
\caption{Impact of the projected EIC SIDIS data on the the sea-quark helicity distributions. 
Together with the DSSV14 estimate, we show the uncertainty bands resulting from the fit that includes the $\sqrt{s}=44.7$ GeV DIS pseudodata, and the reweighting with SIDIS pseudodata at $\sqrt{s}=44.7$ and $\sqrt{s}=141.4$ GeV, respectively.
\label{flavors_refit}}
\end{figure*}

\subsection{Impact of Semi-Inclusive Deep Inelastic Scattering Data}\label{subsec:SIDIS}

In the following we discuss the impact that the EIC measurements of the semi-inclusive production of charged pions and kaons in collisions between longitudinally polarized electrons and protons will have in constraining helicity of quarks.

We restrict the analysis to transverse-momentum integrated final-state hadrons produced in the current-fragmentation region. Even though the QCD framework to describe transverse-momentum–dependent final-state hadron production is known at NLO accuracy 
\cite{Daleo:2004pn} as well as hadron production in the target fragmentation region in terms of fracture functions \cite{Daleo:2003jf,Daleo:2003xg} in the unpolarized case, the helicity dependent framework is still in development.

As we have already shown in Sec.~\ref{subsec:preview}, that charged pion and kaon SIDIS spin asymmetries have the potential to pin down sea quark helicities, complementing inclusive DIS measurements, that at least in the electromagnetic case, are unable to disentangle quark and antiquark helicities. Even though the NLO framework for longitudinally polarized DIS processes mediated by weak vector bosons is well known \cite{deFlorian:1994wp}, it has not been explored yet, leaving pion and kaon SIDIS as the main tools to probe sea quark polarization as a function of the parton momentum fraction. The EIC allows to extend the 
kinematical reach of those measurements and improve dramatically their precision.

In Fig.~\ref{flavors_refit} we show the impact of the projected SIDIS measurements on the sea quark helicity distributions. The light cyan bands in the lefthand, center and righthand side panels represent the uncertainty estimates from DSSV14 for $\Delta \overline{u}$, $\Delta \overline{d}$ and $\Delta \overline{s}$, respectively at $Q^2=10$ GeV$^2$. In the DSSV14 analysis, these distributions are constrained by charged pion, kaon, and unidentified charged hadrons SIDIS data that reach down to 
$x=5.2\,10^{-3}$. In the case of strange quarks, the charge conjugation symmetry assumption, $\Delta s = \Delta \overline{s}$, together with the hyperon semi-leptonic $\beta$-decay data on the full moments, constrain further the helicity distribution.
The sky blue bands in Fig.~\ref{flavors_refit} show the uncertainty estimated by the Monte Carlo sampling variant of DSSV14 that includes also EIC inclusive DIS pseudodata at $\sqrt{s}=44.7$ GeV. An inclusive DIS data set by itself would be unable to constrain the sea-quark densities, however, combined with the SIDIS data already present in the fit, improve the determination in the region of overlap. This effect is milder for $\Delta \overline{u}$ and $\Delta \overline{d}$, however in the case of the strange quarks, the impact is much more noticeable, specially at intermediate and large $x$. Remember that the charge conjugation 
symmetry assumption turns the strange quark distribution effectively into a $\Delta q + \Delta \overline{q}$ quantity for inclusive measurements. On the other hand, the increased flexibility of the new replica set, bypass the hyperon decay constraints at very small-$x$.

The results of reweighting the replicas with EIC SIDIS pseudo-data at $\sqrt{s}=44.7$ and $\sqrt{s}=141.4$ GeV are shown as royal blue and dark blue bands, respectively. The reduction of the uncertainties driven by the inclusion of the SIDIS pseudo-data is much more significant for the three sea-quark distributions 
at $x < 0.01$  As usual, $u$ quarks are much better constrained than $d$ quarks, because of the charge factor. High c.m.s. energy SIDIS data reach smaller values of parton momentum fraction, extending the impact on the bands to much smaller values of momentum fraction. For the strange quarks, the constraint combines a more stringent effect due to the charge symmetry assumption, with a relatively larger uncertainty in the kaon fragmentation function compared to those for pions, that dilutes the effect of the kaon SIDIS asymmetries. 
  
Given the sharp discriminating power of SIDIS for u and d antiquarks, shownin Fig.14, it would be convenient to release the standard assumption on strange polarization $\Delta s = \Delta \overline{s}$, anticipating that if there is a difference between the polarization of strange quarks and antiquarks, it could be in principle pinned down through the comparison of positively and negatively charged
kaon SIDIS measurements at EIC. This is currently beyond the scope of the present exercise, but should be implemented in future global analysis as well as in a quantitative impact assessment.

Another significant advantage of the availability of charge and flavor discriminating SIDIS data, is the possibility of performing the full flavor separation without relaying on hyperon semi-leptonic $\beta$-decay constraints and the flavor symmetry assumptions implicit in their implementation. Global analyses of helicity PDFs routinely use constraints that can be derived from baryonic semi-leptonic $\beta$-decays under the assumption of SU(2) and SU(3) flavor symmetries \cite{ratcl}. These relate combinations of the first moments of the quark helicities to the $F$ and $D$ constants parameterizing the $\beta$-decay rates. However, the use of the beta decay data in this context is controversial; on the one hand because the estimated uncertainties of $F$ and $D$ may not fully reflect the 
actual breaking of the SU(2) and, in particular, SU(3) symmetries, for which larger breaking effects have  been discussed in the literature \cite{su3}. On the other hand, these constraints involve an integration of the helicity distributions over the entire range of momentum fraction, which can not be accessed experimentally, and necessarily involve some kind of extrapolation.
 
In the DSSV analyses rather than imposing the exact SU(2) and SU(3) flavor symmetry relations, deviations are allowed and are quantified in terms of two additional free parameters $\varepsilon_{{\mathrm{SU}}(2)}$ and  $\varepsilon_{{\mathrm{SU}}(3)}$, related to the quark moments and the $F$ and $D$ parameters through:
\begin{eqnarray}
\label{eq:su2n}
\Delta \Sigma_u^{1} - \Delta \Sigma_d^{1} &=&
(F+D)\, [1+\varepsilon_{{\mathrm{SU}}(2)}], \\ \nonumber \\
\label{eq:su3n}
\Delta \Sigma_u^{1} + \Delta \Sigma_d^{1}-2 \Delta \Sigma_s^{1} &=&
(3F-D)\, [1+\varepsilon_{{\mathrm{SU}}(3)}],
\end{eqnarray}
where
\begin{eqnarray}
\label{eq:firstmom1}
\Delta \Sigma_f^{1} 
 \equiv  \int_0^1 \left[ \Delta f_i + \Delta \bar{f}_i \right] (x,\mu_0)\,dx 
\,.
\end{eqnarray}
Note that the free fit parameters $\varepsilon_{{\mathrm{SU}}(2)}$ and $\varepsilon_{{\mathrm{SU}}(3)}$, take the place of the normalization of the $\Delta u+ \Delta \overline{u}$  and $\Delta d+ \Delta \overline{d}$ quark distributions, which otherwise could have been fixed by the measured values of $F$ and $D$. In the analysis, the two combinations including the $F$ and $D$ constants in Eqs.(\ref{eq:su2n}) and (\ref{eq:su3n}),  $F + D$ and $3F - D$  respectively, are taken as two additional data points, i.e., 
are included in the effective $\chi^2$ function, with their corresponding uncertainties \cite{ref:hermes-a1pd} are treated as any other measurement when performing the minimizations. 
The results for the parameters $\varepsilon_{{\mathrm{SU}}(2)}$ and $\varepsilon_{{\mathrm{SU}}(3)}$ obtained optimizing the fit to data represent the degree of symmetry breaking favored by the data included in the fit, or by the statistically equivalent replicas of the data set in the MC approach.

\begin{figure}[t!]
\vspace*{0.05cm}
\begin{center}
\hspace*{-0.1cm}
\epsfig{figure=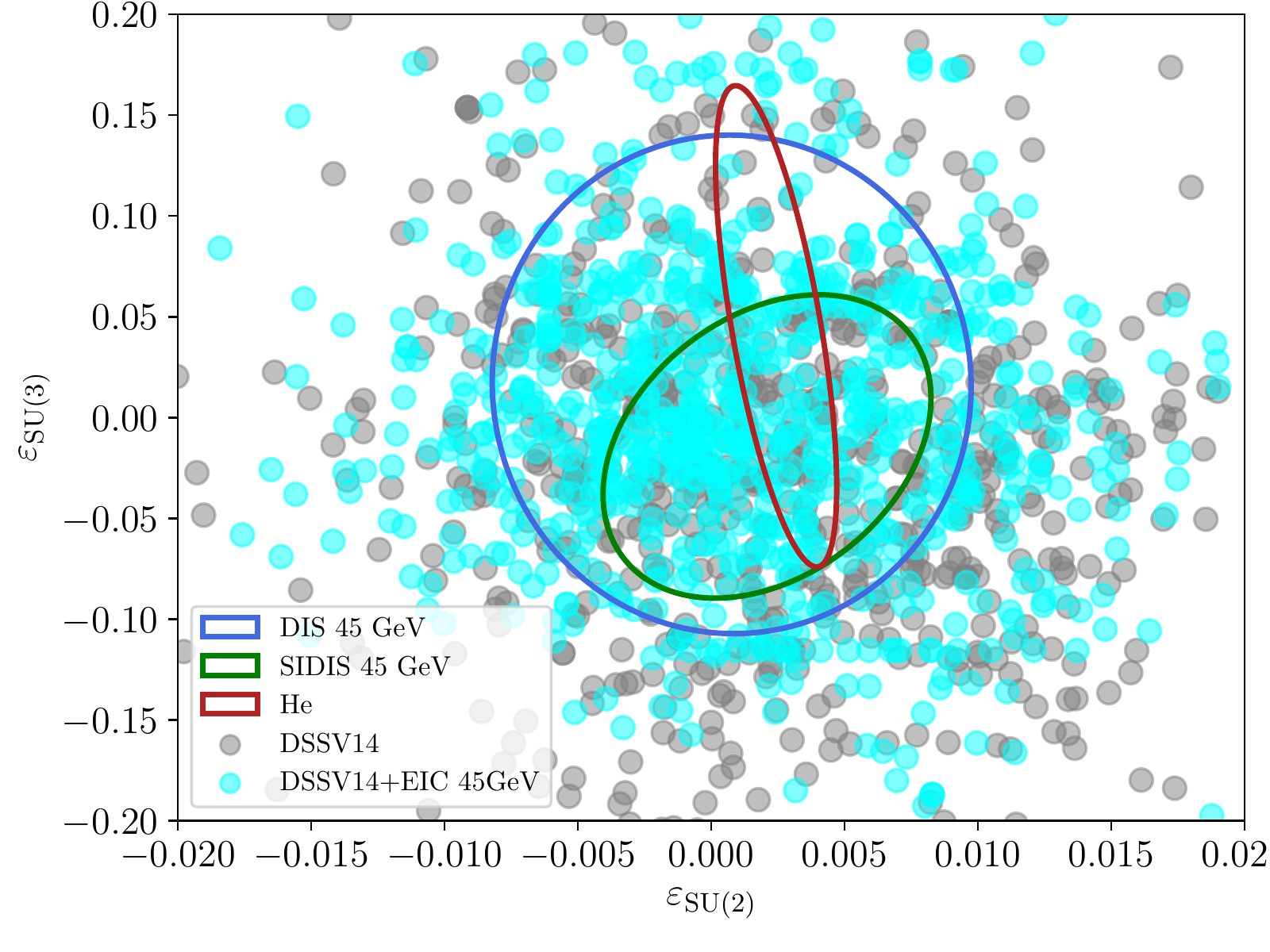,width=0.50\textwidth}
\end{center}
\vspace*{-0.5cm}
\caption{Values for the $\varepsilon_{{\mathrm{SU}}(2)}$ and $\varepsilon_{{\mathrm{SU}}(3)}$ symmetry breaking parameters defined in Eqs.(\ref{eq:su2n}) and (\ref{eq:su3n}), respectively, fitted for the 
DSSV14 MC replicas (grey dots), and those obtained adding also EIC inclusive DIS proton pseudo-data at $\sqrt{s}$ = 44.7 at GeV (cyan), respectively. The contour lines represent the effect of reweighting with SIDIS pseudo-data and inclusive Helium measurements in the 1-$\sigma$ limits.\label{epsilons}}
\end{figure}
In Fig.~\ref{epsilons} we show the values obtained for $\varepsilon_{{\mathrm{SU}}(2)}$ and $\varepsilon_{{\mathrm{SU}}(3)}$ for the DSSV14  MC replicas (grey dots), and those obtained adding also EIC inclusive DIS proton pseudo-data at $\sqrt{s}=44.7$ at GeV (cyan), respectively. Both replica
sets cover a whole range of symmetry breaking values, and are mostly concentrated at $|\varepsilon_{{\mathrm{SU}}(2)}| < 0.01 $ and 
$|\varepsilon_{{\mathrm{SU}}(3)}| < 0.15$, respectively. 
As more data with flavor constraining power are included in the analysis, the symmetry breaking parameters should become better determined. This is precisely what happens when EIC SIDIS pseudo data at $\sqrt{s}=44.7$ at GeV are included through reweighting. In Fig.~\ref{epsilons} 
the green contour represents the 1-$\sigma$ limits for the values of $\varepsilon_{{\mathrm{SU}}(2)}$ and $\varepsilon_{{\mathrm{SU}}(3)}$ after reweighting with SIDIS pseudo-data, to be 
compared with the blue contour, corresponding to the DSSV14 data set plus EIC inclusive DIS proton pseudo-data at $\sqrt{s}=44.7$ at GeV. The addition of SIDIS data reduces roughly by a factor of two the uncertainty on the SU(3) breaking parameter and also limits sizeable the ones for isospin breaking.  Anticipating the results presented in the next section on the EIC He-3 DIS pseudo-data,  we include in Fig.~\ref{epsilons}  the 1-$\sigma$ limits resulting from the corresponding reweighting, as a red contour. As expected, He-3 pseudo-data have a very significant effect in probing isospin symmetry, but make no further improvement in constraining the SU(3) symmetry breaking. This highlights the unique importance of SIDIS as a probe of SU(3) symmetry, through its access to the strange quark helicity, and its complementarity to the inclusive data sets. It also suggests that with the expected level of precision and flavor discrimination, the $\beta$-decay constraints will eventually become not competitive with the
combined effect of EIC DIS and SIDIS data.

\begin{figure*}[t!]
\vspace*{0.05cm}
\begin{center}
\hspace*{-0.1cm}
\epsfig{figure=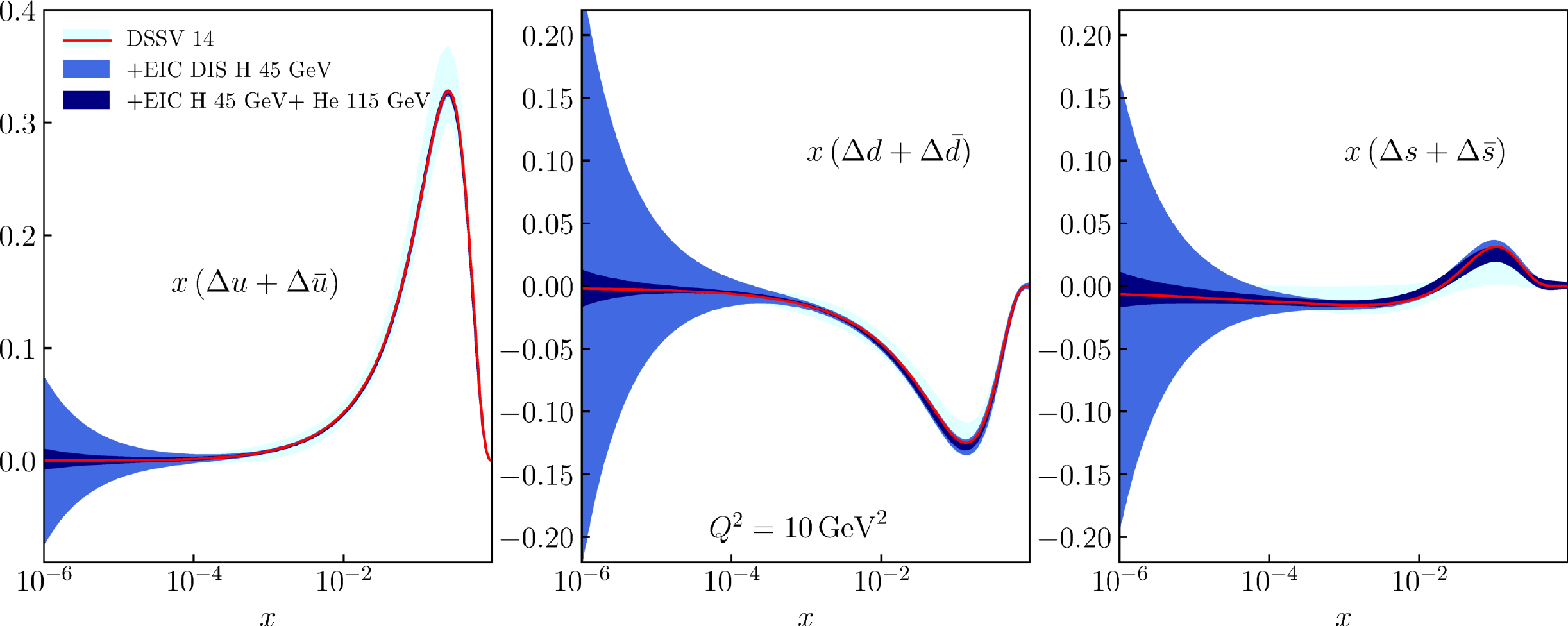,width=0.99\textwidth}
\end{center}
\vspace*{-0.5cm}
\caption{Impact of the projected EIC DIS He data on the total quark helicity distributions $\Delta u + \Delta \overline{d}$, $\Delta d + \Delta \overline{d}$ and $\Delta s + \Delta \overline{s}$, respectively. Together with the DSSV14 estimate, we show the uncertainty bands resulting from the fit that includes the $\sqrt{s}=44.7$ GeV DIS pseudo-data, and the reweighting  with He data at $\sqrt{s}=115.2$ GeV.  \label{dist_He_refit}}
\end{figure*}

\subsection{Impact of Deep Inelastic Scattering off Helium}\label{subsec:He}

Even though SIDIS data achieve a remarkable degree of flavor separation, crucially needed for probing sea quarks, the identification of a final state hadron is always experimentally challenging, and on the other hand, the analysis of such data necessarily involves uncertainties associated to the fragmentation process. In this respect, the availability of polarized light ion beams at EIC, such as deuterium or helium, would allow to have access to a more direct flavor separation for the combined quark plus antiquark 
helicity distributions, but with the precision characteristic of inclusive measurements. This is specially the case for the very low-$x$ regime, where the flavor separation depends exclusively on the scale dependence when using only proton beams, and for $d$ quarks in general, since they are typically relegated to a second rank because of both their electric charge factor and their suppression relative to $u$ quarks in the proton.  

The impact of the EIC pseudo-data on electron-helium collisions at $\sqrt{s}=115.2$ GeV, estimated by reweighting the DSSV14+EIC45 replicas is indeed so significant that even though we started with about a thousand replicas, the number of effective replicas that remain with non-negligible weights is extremely reduced, of order of a dozen replicas, strongly suggesting the need of a new fit. 
Nevertheless, in the following we show the corresponding results, which must be taken as very rough approximation, until a new combined analysis is performed.

\begin{figure}[b!]
\vspace*{0.1cm}
\begin{center}
\hspace*{-0.1cm}
\epsfig{figure=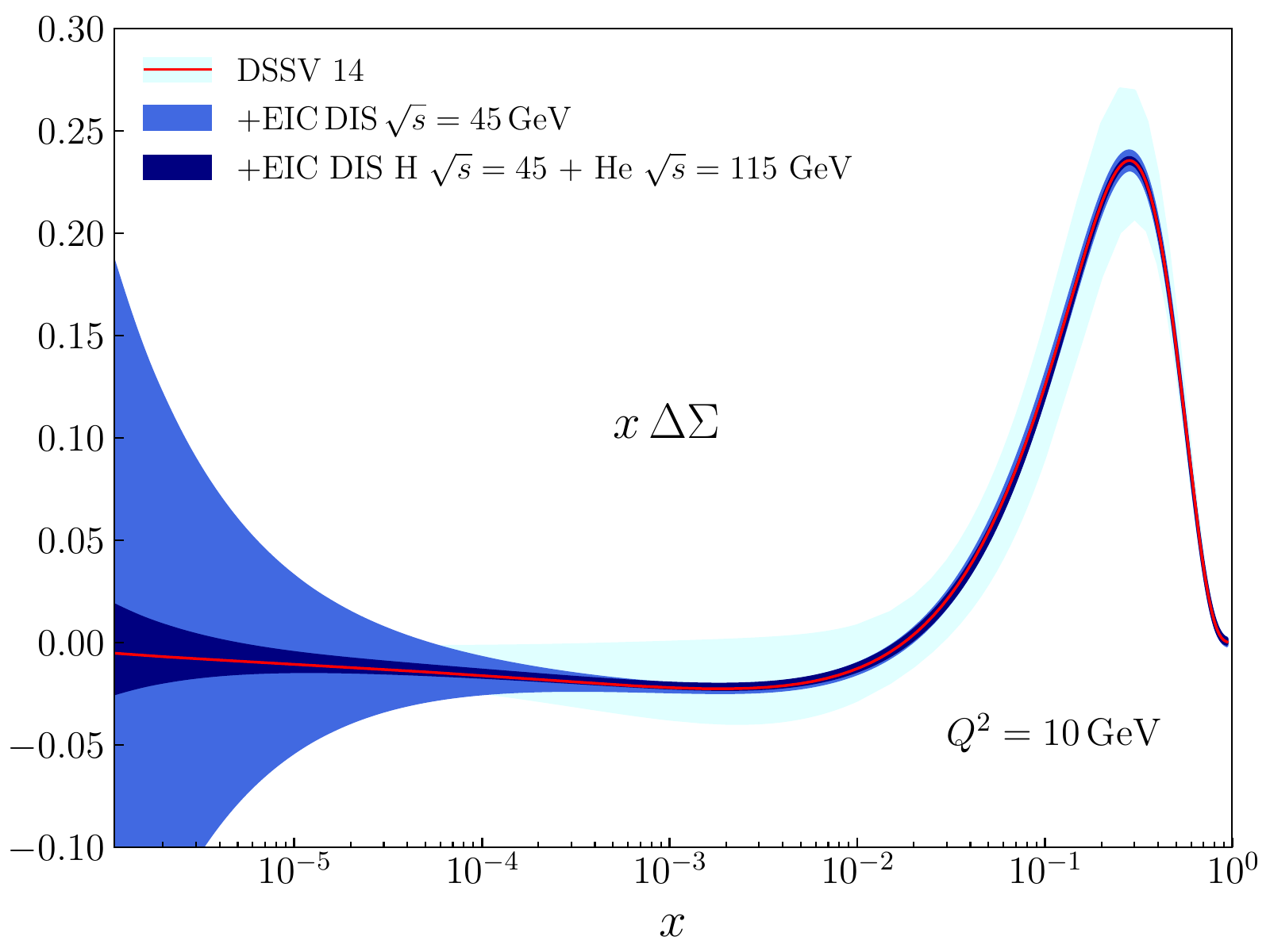,width=0.50\textwidth}
\end{center}
\vspace*{-0.5cm}
\caption{The same as Fig.~\ref{dist_He_refit} but for the singlet quark helicity $\Delta \Sigma$  \label{sigma_He_refit}}
\end{figure}

In Fig.~\ref{dist_He_refit} we show the impact of pseudo-data on inclusive electron-helium collisions at $\sqrt{s}=115.2$ GeV on the total quark helicity distributions $\Delta u + \Delta \overline{u}$, $\Delta d + \Delta \overline{d}$ and $\Delta s + \Delta \overline{s}$, respectively. Notice the rather dramatic reduction in the uncertainty bands at very low $x$, due to the more direct flavor separation. The impact of electron-helium pseudo-data on $\Delta \Sigma$ at small $x$ is also quite significant, as can be seen in Fig.~\ref{sigma_He_refit}, where the estimates for $\Delta \Sigma$ are presented in the full range of parton momentum fraction. 

\begin{figure*}[t]
\vspace*{0.05cm}
\begin{center}
\hspace*{-0.1cm}
\epsfig{figure=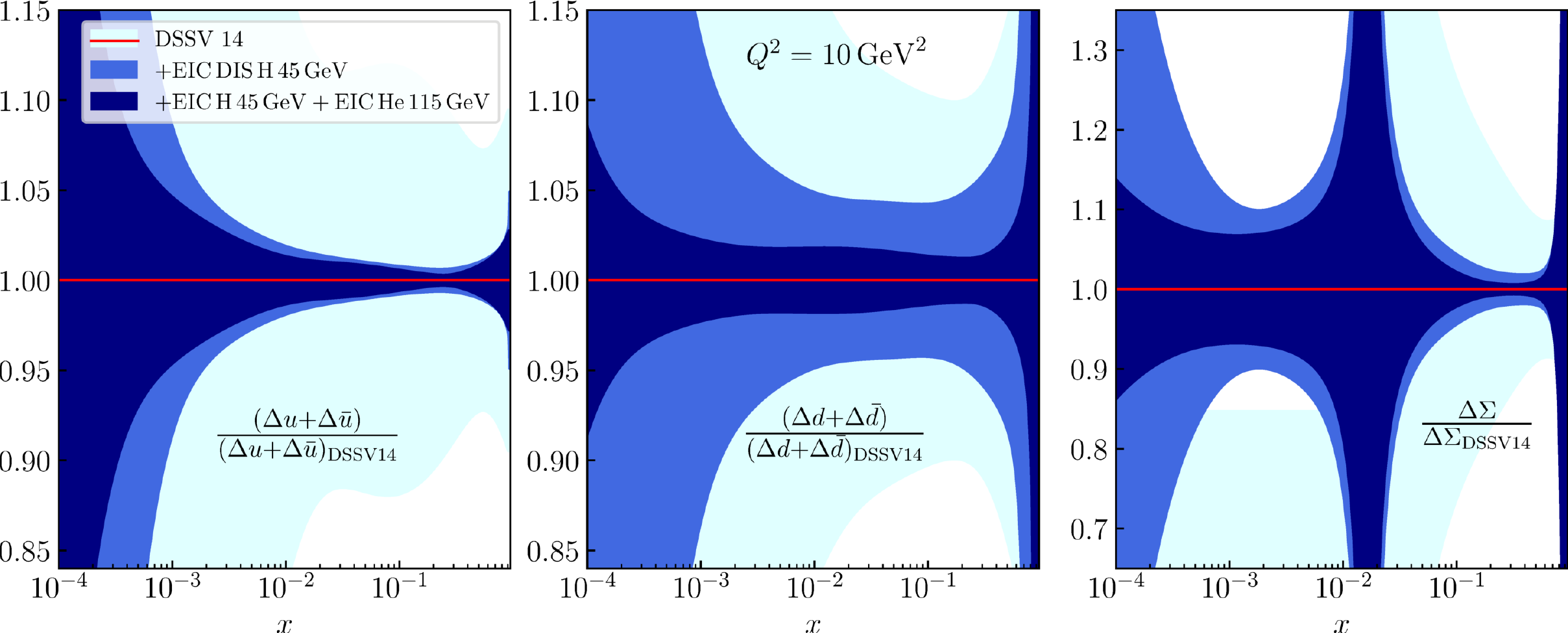,width=0.99\textwidth}
\end{center}
\vspace*{-0.5cm}
\caption{The same as Fig.~\ref{dist_He_refit} but for $\Delta u + \Delta \overline{u}$, $\Delta d + \Delta \overline{d}$ and 
$\Delta \Sigma$ respectively, and presented as ratios to the DSSV14 central value in order to show the relative uncertainties of these distributions.\label{R_He}}
\end{figure*}

It is worth mentioning that strenuous efforts have been made in the last 25 years to compute the small x asymptotic behaviour of helicity distributions within different approaches and formalisms, starting with the pioneering work in Refs. \cite{Bartels:1995iu,Bartels:1996wc} to the most recent results \cite{Kovchegov:2020hgb,Kovchegov:2015pbl,Kovchegov:2016weo,Kovchegov:2017jxc}. The extraction 
of $\Delta \Sigma, \Delta g$  and the individual flavors in the range of momentum fractions $x \sim 10^{-5}$ - $10^{-3}$ would allow to discriminate between the different approaches and approximations, and
help to determine the onset of these very interesting low-$x$ evolution effects.

The impact from the inclusive DIS electron - He-3 data is also very significant for $\Delta d + \Delta \overline{d}$ and $\Delta \Sigma$ at larger values of momentum fractions, as shown in Fig.~\ref{R_He}, where the distributions are normalized to estimates based on the DSSV14 central values.
Notice that in the DSSV14 extraction, the sum $\Delta d + \Delta \overline{d}$ is at best constrained at a 10\% level, in a very narrow range of $x$ in the valence region, as shown by the light cyan band in the center panel of Fig.~\ref{R_He}. The addition of electron-proton EIC pseudo-data at $\sqrt{s}=44.7$ GeV, reduces it to a 5\% level, but only in the region where EIC pseudo-data overlap with flavor sensitive measurements already present in the DSSV14 data set. EIC He pseudo-data constrains $d$ quark 
polarization to a $\sim$ 2\% level down to $x \sim 0.001$, reaching similar precision as for the $u$ quarks, shown in the left panel of Fig.~\ref{R_He}. The pseudo-data also have a significant impact in the 
flavor singlet quark helicity, shown normalized to the estimate from the DSSV14 central value in right panel of Fig.~\ref{R_He}. Since the singlet distribution changes sign at $x\sim 0.02$, the ratio is ill defined around that point, but in any case the reduction in the uncertainty is noteworthy, specially at $x>0.15$ where the distribution approaches its maximun. 

\begin{figure}[t!]
\vspace*{0.1cm}
\begin{center}
\hspace*{-0.1cm}
\epsfig{figure=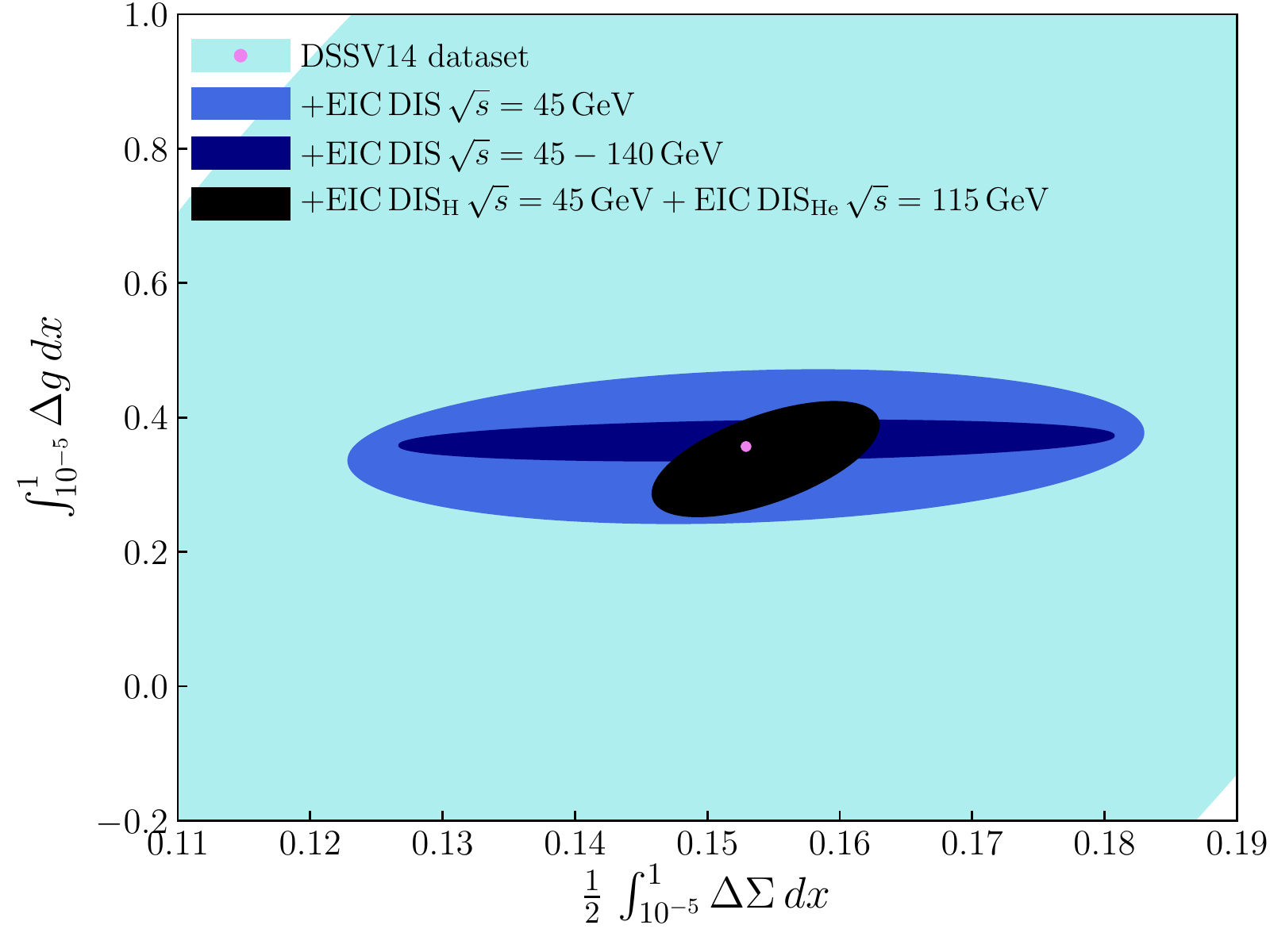,width=0.49\textwidth}
\end{center}
\vspace*{-0.5cm}
\caption{Correlated uncertainties for the truncated integrals of $\Delta g$ and $\Delta \Sigma$ in the range $10^{-5}<x<1$ 
 computed with DSSV14 (outer area), as well as including pseudodata from the EIC (inner areas).  \label{potato_He_refit}}
\end{figure}

Finally, the impact on the flavor singlet quark helicity naturally modifies the uncertainty estimate for its truncated moment, which represents the net contribution of the spin of the quark to proton spin. In Sec.~\ref{subsec:DIS} we showed that its uncertainty was significantly reduced by electron-proton EIC pseudo-data at $\sqrt{s}=44.7$ GeV, but the addition of $\sqrt{s}=141.4$ GeV pseudo-data had a mild effect precisely because of the very limited flavor separation. This would not be the case for electron-helium data, as shown in the correlation plot for the truncated moments down to $x=10^{-5}$. The addition pseudo-data on electron-helium DIS reduces roughly by more than a factor of three the uncertainty on the net quark contribution to the proton spin, as shown in Fig.~\ref{potato_He_refit}.

As it could have been expected that the impact of the electron-helium pseudo-data to the gluon contribution to the proton spin is smaller than that of the $\sqrt{s}=141.4$ GeV inclusive electron-proton pseudo-data mainly because of the more restrictive kinematical coverage of the former. Since the number of surviving replicas obtained combining by reweighting the three pseudo-data sets, namely electron-proton DIS at $\sqrt{s}=44.7$ GeV, $\sqrt{s}=141.4$ GeV and electron-helium DIS at $\sqrt{s}=115.2$ GeV is extremely 
low, we defer a detailed discussion of their combined impact to a new global analysis. Presumably, the combined effect would be at least comparable to the intersection of the two darker areas.

\section{Summary}\label{sec:summary}

We have compared the constraining power on the quark and gluon helicity distributions from inclusive and semi-inclusive electron-proton collisions, and inclusive electron-helium collisions, projected to be measured at the future Electron Ion Collider.
Combining world data on spin dependent processes included in the DSSV14 analysis with updated estimates for the inclusive electron-proton measurements expected to be obtained in a first stage of the EIC, we built a large set of fit replicas using a Monte Carlo sampling approach, and computed the resulting uncertainties. Reweighting the above mentioned replicas with additional pseudo-data sets from inclusive electron-proton DIS at a higher c.m.s. energies, on semi-inclusive production of pions and kaons, and on inclusive electron-helium collisions, we have assessed and compared the impact on the different helicity distributions.

The overall picture we get is one where the first stage of the EIC with electron-proton collisions at $\sqrt{s}=44.7$ GeV reduces the uncertainties of the quark and gluon helicities roughly by a factor 
of four or more down to parton momentum fractions of one thousand of the proton momentum compared to the DSSV14 estimates. Inclusive data constrain mostly the gluon helicity and the total 
flavor singlet quark helicity, while semi-inclusive measurements are indispensable to probe the sea quark helicites. A second stage of the EIC with inclusive collisions at $\sqrt{s}=141.4$ GeV roughly triples the constraining power on the gluons thanks to the increased range in $x$ and $Q^2$, while the semi-inclusive data at this c.m.s. energy, improves the sea quark determination down to much smaller values of momentum fractions. Electron-helium collisions, on the other hand, improve the flavor separation for 
quarks to an unprecedented level, that pushes our reweighting approach to its limits, and suggests the 
need of a full new fit.  

The EIC is designed to bring to conclusion more than forty years of efforts aimed to elucidate the relation between the proton spin and that of its fundamental constituents. This endeavor not only includes the assessment of the role of the gluon spin polarization, and the orbital angular momentum of quarks and gluons, but more in general the relation between spin and flavor, which is at the base of the quark-parton model picture. On the other hand, it also addresses the question about how much of the proton spin depends on those of quarks and gluons carrying negligibly small fractions of the proton momentum. We have investigated to which extent the projected measurements will be able to answer these compelling questions.

\section{Acknowledgments}

I. Borsa and R. Sassot acknowledge Brookhaven National Laboratory for its hospitality and support during the early stages of this work. This work was supported in part by CONICET and ANPCyT. E.C.A and A.S.N acknowledge support from the U.S. Department of Energy under contract number de-sc0012704.



\begin{thebibliography}{99}

\bibitem{ref:emc-a1p}
  J.~Ashman {\it et al.} [European Muon Collaboration],
  Nucl.\ Phys.\  B {\bf 328}, 1 (1989).

\bibitem{Aidala:2012mv} For a review see, e.g.,
  C.~A.~Aidala, S.~D.~Bass, D.~Hasch, and G.~K.~Mallot,
  Rev.\ Mod.\ Phys.\  {\bf 85}, 655 (2013).

\bibitem{Aschenauer:2015eha} 
  E.~C.~Aschenauer {\it et al.},
  {\tt arXiv:1501.01220 [nucl-ex]}.

\bibitem{deFlorian:2014yva}
D.~de Florian, R.~Sassot, M.~Stratmann and W.~Vogelsang,
Phys. Rev. Lett. \textbf{113}, no.1, 012001 (2014).

\bibitem{Nocera:2014gqa} 
  E.~R.~Nocera {\it et al.} [NNPDF Collaboration],
  Nucl.\ Phys.\ B {\bf 887}, 276 (2014).

\bibitem{Accardi:2012qut} 
  A.~Accardi {\it et al.},
  Eur.\ Phys.\ J.\ A {\bf 52}, 268 (2016).

\bibitem{Aschenauer:2019kzf}
E.~C.~Aschenauer, I.~Borsa, R.~Sassot and C.~Van Hulse,
Phys. Rev. D \textbf{99}, no.9, 094004 (2019).

\bibitem{Aschenauer:2015ata}
E.~C.~Aschenauer, R.~Sassot and M.~Stratmann,
Phys. Rev. D \textbf{92}, no.9, 094030 (2015).

\bibitem{Aschenauer:2012ve}
E.~C.~Aschenauer, R.~Sassot and M.~Stratmann,
Phys. Rev. D \textbf{86}, 054020 (2012).

\bibitem{deFlorian:2008mr}
D.~de Florian, R.~Sassot, M.~Stratmann and W.~Vogelsang,
Phys. Rev. Lett. \textbf{101}, 072001 (2008).

\bibitem{deFlorian:2009vb} 
  D.~de Florian, R.~Sassot, M.~Stratmann, and W.~Vogelsang,
  Phys.\ Rev.\ D {\bf 80}, 034030 (2009).

\bibitem{deFlorian:2019zkl}
D.~De Florian, G.~A.~Lucero, R.~Sassot, M.~Stratmann and W.~Vogelsang,
Phys. Rev. D \textbf{100}, no.11, 114027 (2019).

\bibitem{Mankiewicz:1991dp} 
  L.~Mankiewicz, A.~Schafer, and M.~Veltri,
  Comput.\ Phys.\ Commun.\  {\bf 71}, 305 (1992).
  
 \bibitem{ref:qed}
  A.~Kwiatkowski, H.~Spiesberger, and H.~J.~Mohring,
  Comput.\ Phys.\ Commun.\  {\bf 69}, 155 (1992);
  K.~Charchula, G.~A.~Schuler, and H.~Spiesberger,
  Comput.\ Phys.\ Commun.\  {\bf 81}, 381 (1994);
  A.~Arbuzov, D.~Y.~.Bardin, J.~Blumlein, L.~Kalinovskaya, and T.~Riemann,
  Comput.\ Phys.\ Commun.\  {\bf 94}, 128 (1996).   
  
 \bibitem{Sjostrand:2014zea}
    T.~Sjöstrand, et al.,  Comput.\ Phys.\ Commun.\  {\bf 191}, 159 (2015).

\bibitem{deFlorian:2014xna}
D.~de Florian, R.~Sassot, M.~Epele, R.~J.~Hernández-Pinto and M.~Stratmann,
Phys. Rev. D \textbf{91}, no.1, 014035 (2015)
doi:10.1103/PhysRevD.91.014035.

\bibitem{deFlorian:2017lwf}
D.~de Florian, M.~Epele, R.~J.~Hernandez-Pinto, R.~Sassot and M.~Stratmann,
Phys. Rev. D \textbf{95}, no.9, 094019 (2017)
doi:10.1103/PhysRevD.95.094019.

\bibitem{Guffanti:2010yu} 
  A.~Guffanti and J.~Rojo,
  Nuovo Cim.\ C {\bf 033}, no. 4, 65 (2010)
  doi:10.1393/ncc/i2010-10668-y
  [arXiv:1008.4671 [hep-ph]].

\bibitem{Wang:2018heo} 
  B.~T.~Wang, T.~J.~Hobbs, S.~Doyle, J.~Gao, T.~J.~Hou, P.~M.~Nadolsky and F.~I.~Olness,
  arXiv:1803.02777 [hep-ph].
 
\bibitem{DelDebbio:2007ee} 
  L.~Del Debbio {\it et al.} [NNPDF Collaboration],
  JHEP {\bf 0703}, 039 (2007).
\bibitem{Ball:2008by} 
  R.~D.~Ball {\it et al.} [NNPDF Collaboration],
  Nucl.\ Phys.\ B {\bf 809}, 1 (2009),
  Erratum: [Nucl.\ Phys.\ B {\bf 816}, 293 (2009)].

\bibitem{Stump:2001gu} 
  D.~Stump, J.~Pumplin, R.~Brock, D.~Casey, J.~Huston, J.~Kalk, H.~L.~Lai, and W.~K.~Tung,
  Phys.\ Rev.\ D {\bf 65}, 014012 (2001).
  
\bibitem{Pumplin:2001ct} 
  J.~Pumplin, D.~Stump, R.~Brock, D.~Casey, J.~Huston, J.~Kalk, H.~L.~Lai, and W.~K.~Tung,
  Phys.\ Rev.\ D {\bf 65}, 014013 (2001).

\bibitem{Ball:2010gb} 
  R.~D.~Ball {\it et al.} [NNPDF Collaboration],
  Nucl.\ Phys.\ B {\bf 849}, 112 (2011),
  Erratum: [Nucl.\ Phys.\ B {\bf 854}, 926 (2012)],
  Erratum: [Nucl.\ Phys.\ B {\bf 855}, 927 (2012)].
%
\bibitem{Ball:2011gg} 
  R.~D.~Ball {\it et al.},
  Nucl.\ Phys.\ B {\bf 855}, 608 (2012).

\bibitem{Armesto:2013kqa} 
  N.~Armesto, J.~Rojo, C.~A.~Salgado, and P.~Zurita,
  JHEP {\bf 1311}, 015 (2013).
%
\bibitem{Paukkunen:2014zia} 
  H.~Paukkunen and P.~Zurita,
  JHEP {\bf 1412}, 100 (2014).
%
\bibitem{Borsa:2017vwy} 
  I.~Borsa, R.~Sassot, and M.~Stratmann,
  Phys.\ Rev.\ D {\bf 96}, 094020 (2017).

\bibitem{Stratmann:2001pb} 
  M.~Stratmann and W.~Vogelsang,
  Phys.\ Rev.\ D {\bf 64}, 114007 (2001).

\bibitem{ap1} R.~Mertig and W.~L.~van Neerven, Z.\ Phys.\  C {\bf 70}, 637 (1996);
  W.~Vogelsang, Phys.\ Rev.\  D {\bf 54}, 2023 (1996); Nucl.\ Phys.\  B {\bf 475}, 
  47 (1996).

\bibitem{deFlorian:1994wp}
D.~de Florian and R.~Sassot,
Phys. Rev. D \textbf{51}, 6052-6058 (1995)

\bibitem{Daleo:2004pn}
A.~Daleo, D.~de Florian and R.~Sassot,
Phys. Rev. D \textbf{71}, 034013 (2005)

\bibitem{Daleo:2003jf}
A.~Daleo and R.~Sassot,
Nucl. Phys. B \textbf{673}, 357-384 (2003)

\bibitem{Daleo:2003xg}
A.~Daleo, C.~Garcia Canal and R.~Sassot,
Nucl. Phys. B \textbf{662}, 334-358 (2003)

\bibitem{ratcl} For review, see: P.~G.~Ratcliffe,
  Czech.\ J.\ Phys.\  {\bf 54}, B11 (2004).

\bibitem{su3}  H.~J.~Lipkin,
  Phys.\ Lett.\  B {\bf 214}, 429 (1988);
  Phys.\ Lett.\  B {\bf 230}, 135 (1989);
  F.~E.~Close and R.~G.~Roberts,
  Phys.\ Rev.\ Lett.\  {\bf 60}, 1471 (1988);
  M.~Roos,
  Phys.\ Lett.\  B {\bf 246}, 179 (1990);
  Z.~Dziembowski and J.~Franklin,
  J.\ Phys.\ G {\bf 17}, 213 (1991).
  P.~G.~Ratcliffe,
  Phys.\ Lett.\  B {\bf 242}, 271 (1990);
  Phys.\ Lett.\  B {\bf 365}, 383 (1996); 
  {\tt arXiv:hep-ph/0012133};
  S.~L.~Zhu, G.~Sacco, and M.~J.~Ramsey-Musolf,
  Phys.\ Rev.\  D {\bf 66}, 034021 (2002);
  E.~Leader and D.~B.~Stamenov,
  Phys.\ Rev.\  D {\bf 67}, 037503 (2003).

\bibitem{ref:hermes-a1pd}
  A.~Airapetian {\it et al.}  [HERMES Collaboration],
  Phys.\ Rev.\  D {\bf 75}, 012007 (2007).

\bibitem{Bartels:1995iu}
J.~Bartels, B.~I.~Ermolaev and M.~G.~Ryskin,
Z. Phys. C \textbf{70}, 273-280 (1996)
\bibitem{Bartels:1996wc}
J.~Bartels, B.~I.~Ermolaev and M.~G.~Ryskin,
Z. Phys. C \textbf{72}, 627-635 (1996)
doi:10.1007/BF02909194

\bibitem{Kovchegov:2020hgb}
Y.~V.~Kovchegov and Y.~Tawabutr,
[arXiv:2005.07285 [hep-ph]].
\bibitem{Kovchegov:2015pbl}
Y.~V.~Kovchegov, D.~Pitonyak and M.~D.~Sievert,
JHEP \textbf{01}, 072 (2016)
doi:10.1007/JHEP01(2016)072
\bibitem{Kovchegov:2016weo}
Y.~V.~Kovchegov, D.~Pitonyak and M.~D.~Sievert,
Phys. Rev. Lett. \textbf{118}, no.5, 052001 (2017)
doi:10.1103/PhysRevLett.118.052001
\bibitem{Kovchegov:2017jxc}
Y.~V.~Kovchegov, D.~Pitonyak and M.~D.~Sievert,
Phys. Lett. B \textbf{772}, 136-140 (2017)
doi:10.1016/j.physletb.2017.06.032


\end{thebibliography}
\end{document}